\documentclass[12pt,twoside]{article} 
\usepackage[sectionbib]{natbib}
\usepackage{array,epsfig,fancyheadings,rotating}
\usepackage[]{hyperref}  
\usepackage{setspace}

\usepackage{sectsty, secdot}
\sectionfont{\fontsize{12}{14pt}\selectfont}
\renewcommand{\theequation}{\thesection\arabic{equation}}
\subsectionfont{\fontsize{12}{14pt plus.8pt minus .6pt}\selectfont} 
\allsectionsfont{\singlespacing}

\usepackage{amsmath}
\usepackage{amssymb}
\usepackage{amsfonts}
\usepackage{multirow}
\usepackage{amsthm}

\newtheorem{theorem}{Theorem}

\theoremstyle{definition}
\newtheorem{definition}{Definition}
\newtheorem{example}{Example}

\usepackage{graphicx}
\usepackage{amssymb}
\usepackage{bm}
\usepackage{arydshln}
\usepackage{amsmath}
\usepackage{caption}
\usepackage{hyperref}
\usepackage{comment}
\usepackage{natbib}
\usepackage{multirow}

\newcommand{\bfzeta}{\mbox{\boldmath $\zeta$}}
\newcommand{\bfzero}{\mbox{\boldmath $0$}}
\newcommand{\bftheta}{\mbox{\boldmath $\theta$}}
\newcommand{\bfmu}{\mbox{\boldmath $\mu$}}
\newcommand{\bfy}{\mbox{\boldmath $y$}}

\begin{document}
\renewcommand{\baselinestretch}{2}
\singlespacing  


\markboth{\hfill{\footnotesize\rm S. SHARIFI FAR, M. PAPATHOMAS AND R. KING} \hfill}
{\hfill {\footnotesize\rm PARAMETER REDUNDANCY IN LOG-LINEAR MODELS} \hfill}



\fontsize{12}{14pt plus.8pt minus .6pt}\selectfont 
\centerline{\large\bf PARAMETER REDUNDANCY AND THE EXISTENCE }
\vspace{2pt} \centerline{\large\bf OF MAXIMUM LIKELIHOOD ESTIMATES }
\vspace{2pt} \centerline{\large\bf  IN LOG-LINEAR MODELS}
\vspace{.4cm} \centerline{Serveh Sharifi Far, Michail Papathomas* and Ruth King} \vspace{.2cm} \centerline{\it
University of Edinburgh and *University of St Andrews} \vspace{.55cm} \fontsize{9}{11.5pt plus.8pt minus .6pt}\selectfont

\thispagestyle{plain}

\begin{quotation}
\noindent {\it Abstract:}
Log-linear models are typically fitted to contingency table data to describe and identify the relationship between different categorical variables. However, the data may  include observed zero cell entries.
The presence of zero cell entries can have an adverse effect on the estimability of parameters, due to  parameter redundancy. We describe a general approach for determining whether a given log-linear model is parameter redundant for a pattern of observed zeros in the table, prior to fitting the model to the data. We derive the estimable parameters or functions of parameters  and also explain how to reduce the unidentifiable model to an identifiable one. Parameter redundant models have a flat ridge in their likelihood function. We further explain when this ridge imposes some additional parameter constraints on the model, which can lead to obtaining  unique maximum likelihood estimates for parameters that otherwise would not have been estimable. In contrast to other frameworks, the  proposed novel approach informs on those constraints, elucidating the model that is actually being fitted.

\vspace{9pt}
\par
\end{quotation}\par


\def\thefigure{\arabic{figure}}
\def\thetable{\arabic{table}}

\renewcommand{\theequation}{\thesection.\arabic{equation}}

\fontsize{12}{14pt plus.8pt minus .6pt}\selectfont

\setcounter{section}{0} 
\setcounter{equation}{0} 


\section{Introduction} 

Observations from multiple categorical random variables
can be cross-classified according to the combinations of the variables' levels. This type of data is often displayed in a contingency table where each cell count is the number of  subjects with a given cross-classification. Log-linear models are typically fitted to such tables  and  examples of their applications  are given by  \citet{Agresti:2002}, \citet{Bishop.etal:1975}  and  \citet{McCullagh:1989}. 

Zero cell counts can have an adverse effect on the estimability of log-linear model parameters. Zero entries are of two main types; structural and sampling zeros. If the expectation and variance of a cell count  are zero, then the entry is a structural zero. A sampling zero is an observed zero entry to a cell with positive expectation. In this manuscript, we examine how zero cell entries influence the estimability of  log-linear model parameters, and this is addressed with respect to parameter redundancy.

A model is not identifiable if two different sets of parameter values generate the same model for the data, which often happens when a model is over-parametrised. This cause of non-identifiability is termed parameter redundancy \citep{Catchpole:1997}. A parameter redundant model can be rearranged as a function of a smaller set of parameters, which are themselves  functions of the initial parameters.  
Parameter redundant models have a flat ridge in their likelihood surface which precludes unique maximum likelihood estimates  for some of the parameters \citep{Catchpole:1997}. For a log-linear parameter redundant model, often undefined or large standard errors  for nonestimable parameters  are reported by numerical optimisation methods. An overview of  identifiability and parameter redundancy is given by  \citet{Catchpole:1997} and \citet{Catchpole:1998}. \citet{Cole:2010} provide several ecological examples on this topic. Identifiability is crucial when exploring  complex associations between factors, as interaction terms quickly become nonestimable in the presence of zero cell counts.  The development of methods that identify the highest level of interaction complexity, which can be explored for a given data set, is therefore important.

We develop a method for the detection of parameter redundancy for log-linear models in the presence of sampling zero observations. The estimable parameters and combinations of parameters are derived, and it is shown how a parameter redundant model can be reduced to a non-redundant one which is also identifiable.  We refer to the proposed method as the ``parameter redundancy'' approach.  In the presence of structural zeros, the corresponding cells are omitted from the modelling and analysis,  since they are associated with cross-classifications that cannot be observed. 

A comprehensive study of log-linear models for contingency tables was developed by \citet{Haberman:1973}, who proved that  maximum likelihood estimates of  model parameters are unique when they exist, and provided a necessary and sufficient condition for the existence of cell mean estimates in the presence of zero cell entries. This was further studied by \citet{Brown:1983} via  considering and comparing iterative methods, and by \citet{Lauritzen:1996} via a polyhedral and graphical model framework.  A polyhedral version of Haberman's  condition for the existence of the maximum likelihood estimator (MLE) is provided by \citet{Eriksson:2006}. Estimability of parameters under a non-existent MLE, within the extended exponential families, is studied by  \citet{Fienberg:2012a}, and is developed to higher dimensional problems by \citet{Wang:2016}. We refer to these developments  collectively as the ``Existence of the Maximum Likelihood Estimator'' or EMLE framework. The method demonstrates that some of the parameters cannot be estimated when the MLE does not exist. However, an extended estimator, where some of the elements of the estimated  cell mean vector are zero, always exists  \citep{Eriksson:2006}.  In this case, it is possible to reduce the model and estimate a subset of the initial parameters.

We compare the proposed parameter redundancy approach with the EMLE method. The reduced models obtained by the two methods may differ in terms of their parametrisation, but the parameter redundancy approach provides a  reparametrisation that retains the original interpretation of the parameters. This is because this method provides estimable parameters and linear combinations of parameters instead of just the estimable subset of the model's initial parameters.  The parameter redundancy approach also reveals additional constraints imposed by the likelihood function on some parameter redundant models. 
Standard statistical software packages report parameter estimates for such a model without informing on the additional implied constraints. 

Section \ref{sec:loglin} introduces the necessary notation.  Section \ref{sec:parred} describes the determination of a parameter redundant model and the proposed adaptation to log-linear models. The idea is illustrated by examples and a study on saturated log-linear models. We also show   when additional constraints enable us to determine unique ML estimates for additional parameters, thus specifying the model that is in fact fitted to the sparse table.  In Section \ref{sec:exstMLE},  the EMLE framework is reviewed,  and in Section \ref{sec:comparison}, the two approaches are compared using illustrative examples.  Section \ref{sec:discussion} concludes with a discussion.


\subsection{Log-linear models for contingency tables}\label{sec:loglin}
Adopting the notation in \citet{Overstall:2014}, let $ {V}=\{V_1,\dots,V_m\} $ denote a set of $ m $ categorical variables, where the $ j $th variable has $ l_j $ levels. The corresponding contingency table has $ n=\prod_{j=1}^{m} l_j $ cells. Let $ \textbf{y} $ denote an $ n \times 1 $ vector corresponding to the observed cell counts. Each element of $ \textbf{y} $ is denoted by $ y_\textbf{i}$, $ \textbf{i}=(i_1 \dots i_m)$ such that $ 0\leqslant i_j \leqslant l_j-1$ and $j=1,\dots,m  $. Here, $\textbf{i}$, identifies the combination of variable levels that cross-classify the given cell. We define $ L $ as the set of all $ n $ cross-classifications, so that $ L=\otimes _{j=1} ^{m} [ l_j ] $, in which $ [l_j] =\{0,1,\dots,l_j-1\} $. Then, $ N=\sum_{\textbf{i} \in L} y_\textbf{i}  $ denotes the sum of all cell counts. 
The $ y_{\textbf{i}} $s are assumed to be observations from independent Poisson random variables, $Y_{\textbf{i}}$, 
such that, $ \mu_\textbf{i}={E}(Y_\textbf{i}) $.  Let $ \mathcal{E} $ denote a set of subsets of $V$. By adapting the notation of \citet{Johndrow:2014}, the log-linear model assumes the form,
\begin{equation}\label{loglinear model formula}
m_{\textbf{i}}=\log \mu_\textbf{i} =\sum_{e\in \mathcal{E}} \theta^e(\textbf{i}), 
\end{equation}
where $ {\theta}^e(\textbf{i}) \in \mathcal{R} $ denotes the main effect or the interaction among the variables in $ e $  corresponding to the levels in $ \textbf{i} $. The summation is over all members of $\mathcal{E}$, which could be the set of all subsets of the variables (for a saturated model) or a set of desirable subsets (for a smaller model). As a convention, $ \theta $ corresponds to $ e=\emptyset $, so that when the set $\mathcal{E}$ contains $ e=\emptyset $ there is an intercept $ \theta $ in the model.
To allow for the existence of unique parameter estimates, corner point constraints are applied, so that parameters that incorporate the lowest level of a variable are set to zero.   To clarify the notation, consider this minimal example. Assume two categorical variables, $ V=\{X,Y\} $, with $ l_1=l_2=2 $ levels. Then, the number of cells in the $ l_1\times l_2 $ table is $ 4 $ and $ L=\{00,10,01,11\} $. The set of subsets of  $ V $,  $ \mathcal{E}=\{\emptyset,\{X\}, \{Y\} \} $ constructs   the following independence log-linear model, shown as model ($X,Y$), 
$$ \begin{array}{ll}
m_{00}= \log {\mu}_{00} =\theta, & m_{10}= \log {\mu}_{10}=\theta+\theta^{X}_1, \\ 
m_{01}= \log {\mu}_{01} =\theta+\theta^{Y}_1, & 
m_{11}= \log {\mu}_{11} =\theta+\theta^{X}_1+\theta^{Y}_{1}. 
\end{array} $$

Alternatively to (\ref{loglinear model formula}), for $p$ parameters, we can write,  $ \textbf{m}_{n\times 1}=\log\boldsymbol \mu_{n\times 1}=A_{n \times p}\boldsymbol\theta_{p \times 1} $, where $A$ is a full rank design matrix with elements $ \{0,1\} $. Therefore, this model can be written as below, in which the subscript indices  of parameters are removed because there are only two possible variable levels,

{\small $$\left[\def\arraystretch{0.55}
	\begin{array}{c}
	\log\mu_{00}\\
	\log\mu_{10} \\
	\log\mu_{01} \\
	\log\mu_{11} 
	\end{array}
	\right]= \left[\def\arraystretch{0.55}
	\begin{array}{ccc}
	1 & 0 & 0  \\
	1 & 1 & 0  \\
	1 & 0 & 1  \\
	1 & 1 & 1 
	\end{array}
	\right]\left[\def\arraystretch{0.55}
	\begin{array}{c}
	\theta\\
	\theta^{X}\\
	\theta^{Y}
	\end{array}
	\right]. $$}

For a model fitted to an $l^m$ table (with $ m $ variables, each classified in $ l $ levels), an alternative  notation to denote cell counts in (\ref{loglinear model formula}) is possible by setting a one-to-one correspondence between the elements of  $ L $ and integers, $ i=1,\dots,l^m $, as  
\begin{equation}\label{order}
\textbf{i}=(i_1\dots i_m)=i_1l^0+i_2l^1+ \dots +i_{m-1}l^{m-2}+i_ml^{m-1}+1.
\end{equation}  
Thus, for the mentioned example, elements in $ L=\{00,10,01,11\} $  correspond to $ \{1,2,3,4\}$ respectively.


\section{The Parameter Redundancy approach} \label{sec:parred}

\subsection{The derivative method}\label{method}
\citet{Goodman:1974} first used a derivative approach to detect identifiability in latent structure models and   \textit{m}-way contingency tables.  The generic approach for the exponential family of distributions that we summarize here was presented by \citet{Catchpole:1997} and \citet{Catchpole:1998}, and  was also developed independently by \citet{Chappell:1998} and \citet{Evans:2000} for compartmental models.

The mean vector $  \boldsymbol\mu={E}(\textbf{Y}) $ of observations    from a distribution that belongs to the exponential family of distributions,  is expressible as a function of parameters $ \boldsymbol\theta=(\theta_1,\dots,\theta_p) $. The derivative matrix $D(\boldsymbol\theta)$, which describes the relationship between $ \boldsymbol\mu $ (or a monotonic function of it) and $ \boldsymbol\theta $, has elements,
\begin{equation}\label{Dmat}
D_{si}(\boldsymbol\theta)=\frac{\partial \mu_i}{\partial \theta_s}, \qquad s=1,\dots,p, \quad  i=1,\dots,n. 
\end{equation} 
Theorem 1 of \citet{Catchpole:1997} states that the model which relates  $\boldsymbol\mu$ to $\boldsymbol\theta$ is parameter redundant if and only if the derivative matrix is symbolically rank deficient. That is if there exists a non-zero vector $ \boldsymbol\alpha(\boldsymbol\theta) $ such that  for all $ \bm\theta $,
\begin{equation}\label{alpha}
\boldsymbol\alpha(\boldsymbol\theta)^\mathsf{T}D(\boldsymbol\theta)=\textbf{0}. 
\end{equation} 
As an alternative, \citet{Cole:2010}  construct a derivative matrix by differentiating an ``exhaustive summary'' of the model. An exhaustive summary  is a vector of parameter combinations that uniquely defines the model. 

The rank of the derivative matrix, $ r $, is the number of estimable parameters and  combinations of parameters. The model deficiency is defined as $d=p-r  $, which is the number of linearly independent $ \boldsymbol\alpha(\boldsymbol\theta) $ vectors, labelled as $ \boldsymbol\alpha_j(\boldsymbol\theta), \,j=1,\dots,d. $ Any elements of these vectors which are zero for all $ j $, correspond to the parameters that are directly estimable \citep{Catchpole:1998}.  To find the estimable combinations of parameters, the auxiliary equations of the following system of linear first order partial differential equations need to be solved,
\begin{equation}\label{de}
\sum_{s=1}^{p}\alpha_{sj}\frac{\partial f}{\partial \theta_s}=0, \qquad j=1,\dots,d,
\end{equation} 
\citep{Catchpole:1998}. The solution can be obtained using software such as \verb|Maple| which allows symbolic computations. 


\subsection{Parameter redundancy for log-linear models} \label{parred}

Parameter redundancy  occurs due to the model structure or  lack of data  \citep{Catchpole:2001, Cole:2010}, and the latter  type is  referred to as ``extrinsic'' parameter redundancy \citep{Gimenez:2004}. Model (\ref{loglinear model formula}) is constructed so that it is not over-parametrised due to its structure. To detect extrinsic parameter redundancy for a log-linear model,  we  adjust the derivative matrix elements (\ref{Dmat}) using $y_{i} \log\mu_{i}$  as a monotonic function of $ \mu_i $, such that,
\begin{equation}\label{Dmat ct}
D_{si}=\frac{\partial y_{i} \log\mu_{i}}{\partial \theta_s}, \qquad s=1,\dots,p, \quad  i=1,\dots,n.
\end{equation}
In effect, each sampling zero turns a column of the derivative matrix to zero and may decrease the rank of the derivative matrix. 

If the rank of the derivative matrix is  smaller than $p$, the model is parameter redundant. Finding all estimable parameters and estimable combinations of parameters further identifies which cell means are estimable. The vector of estimable quantities $ (\bm{\theta}')$ and the vector of estimable cell means  $ (\bm{\mu}') $ specify a reduced model via a smaller design matrix $ (A') $.  The reduced model is full rank with rank $r$, and its degrees of freedom is the number of estimable cell means minus $r$. 

To clarify the notation,  consider the independence log-linear model ($ X,Y $) for a $ 2 \times 2 $ table. 
The derivative matrix (\ref{Dmat ct}) for observations $ \textbf{y}^\mathsf{T}=(y_1,y_2,y_3,y_4) \\= (y_{00},y_{10},y_{01},y_{11})$ and parameters  $\bm\theta^\mathsf{T}=(\theta,\theta^X,\theta^Y)$ is,
\begin{align*}
{\small D=\left[ \frac{\partial y_{i} \log\mu_{i}}{\partial   \theta_s}\right] =
	\left[\def\arraystretch{0.7}
	\begin{array}{c|cccc}
	& \mu_{00} & \mu_{10} & \mu_{01} & \mu_{11}  \\
	\hline
	\theta & y_1 & y_2& y_3& y_4  \\
	\theta^X &0  &y_2 &0   &y_4\\  
	\theta^Y &0  &0   &y_3 &y_4
	\end{array}
	\right],  \quad s=1,2,3, \,\, i=1,2,3,4.}
\end{align*}
Now, for example, assume that  $ y_1=y_2=0 $. Then, $ r=2 $, $\, d=1 $ and $ \bm\alpha^\mathsf{T}=(1,0,-1) $. Equation (\ref{de}) is $ \frac{\partial f}{\partial \theta}-\frac{\partial f}{\partial \theta^Y}=0 $ and solving it gives the estimable  parameters  $\bm\theta'^{\mathsf{T}}=(\theta^X,\theta+\theta^Y)$. It determines that only $ \bm{\mu}'^{\mathsf{T}}=(\mu_{01}, \mu_{11}) $ are estimable. Therefore, the reduced design matrix $A' $ is $ 2 \times 2 $ with  two rows $ [(0,1), (1,1)] $.

Alternative approaches for investigating identifiability are not suitable in the context of Poisson log-linear models for contingency tables. Specifically, using the log-likelihood function elements as exhaustive summaries is a common option in forming the derivative matrix \citep{Cole:2010}. Similarly, \citet{Catchpole:2001} use the score vector of a multinomial log-linear model to assess the effect of missing data on the model redundancy. Also, utilizing the information matrix instead of a derivative matrix is  an alternative for detecting non-identifiability  \citep{Rothenberg:1971}. 
However, these approaches  do not necessarily show the rank deficiency caused  by the zero cell counts for a Poisson log-linear model.  
The next two examples further illustrate the use of the parameter redundancy method. 

\begin{example}\label{example1}
	The data pattern in Table \ref{3by3}, taken from \citet{Fienberg:2012a}, describes cell counts for variables $ X $ (rows), $ Y $ (columns), and $ Z $ (layers), with three levels $ (0,1,2) $ for each. Eight cell counts are observed as sampling zeros. All  other cell counts are positive Poisson observations, numbered according to (\ref{order}). We fit the hierarchical model  $ (XY,XZ,YZ) $ which can be shown as  
	$ \log \boldsymbol \mu _{27 \times 1} = A_{27 \times 19}\boldsymbol \theta_{19 \times 1}, $
	with parameters,
	\begin{align*}
	\bm{\theta}^\mathsf{T}= (& \theta, \theta^X_1, \theta^X_2, \theta^Y_1, \theta^Y_2, \theta^Z_1, \theta^Z_2, \theta^{XY}_{11}, \theta^{XY}_{21}, \theta^{XY}_{12}, \theta^{XY}_{22}, \\ &\theta^{YZ}_{11}, \theta^{YZ}_{21}, \theta^{YZ}_{12}, \theta^{YZ}_{22}, \theta^{XZ}_{11}, \theta^{XZ}_{21}, \theta^{XZ}_{12}, \theta^{XZ}_{22}).
	\end{align*}	
	The matrix form of this model  is given in the Supplementary Material.
	\begin{table}[t!]
		\caption{ Observations in a $ 3^3 $ contingency table }\label{3by3}
		\begin{center}
			\begin{normalsize}
				\def\arraystretch{0.9}
				\begin{tabular}{c c c }
					\begin{minipage}{0.16\linewidth}
						\centering	
						\begin{tabular}{|c|c|c|}
							\hline
							0 & $ y_4 $ & $ y_7 $\\
							\hline
							0 & $ y_5 $ & $ y_8 $\\
							\hline
							$ y_3 $ & $ y_6 $ & $ y_9 $\\
							\hline
						\end{tabular} 
					\end{minipage} &
					\begin{minipage}{0.19\linewidth}
						\centering	
						\begin{tabular}{|c|c|c|}
							\hline
							$ y_{10} $ & $ y_{13} $ & $ y_{16} $\\
							\hline
							$ y_{11} $ & $ y_{14} $ & $ 0^* $\\
							\hline
							$ y_{12} $ & 0 & 0\\
							\hline
						\end{tabular} 
					\end{minipage} &
					\begin{minipage}{0.2\linewidth}
						\centering	
						\begin{tabular}{|c|c|c|}
							\hline
							0 & $ y_{22} $ & $ 0^* $\\
							\hline
							0 & $ y_{23} $ &  $ y_{26} $\\
							\hline
							$ y_{21} $ & $ y_{24} $ & $ y_{27} $ \\
							\hline
						\end{tabular} 
					\end{minipage} 
				\end{tabular} 
			\end{normalsize}
		\end{center}
	\end{table}

	The rank of the derivative matrix in accordance with  (\ref{Dmat ct}) is 18,  i.e. there are only 18 estimable parameters or combinations of them. So, $d=19-18=1$, and the $ \boldsymbol{\alpha} $  that satisfies (\ref{alpha}) is, 
	$\boldsymbol{\alpha}^\mathsf{T}=(1,0,-1,-1,-1,-1,
	0,0,1,0,1,1,1,0,0\\,0,1,0,0). $
	Solving (\ref{de}) gives the estimable quantities as,
	\begin{align*}
	\boldsymbol{\theta}'^{\mathsf{T}}= (&\theta^X_{1}, \theta+\theta^X_{2}, \theta+\theta^Y_{1}, \theta+\theta^Y_{2}, \theta+\theta^Z_{1}, \theta^Z_{2}, \theta^{XY}_{11},-\theta+\theta^{XY}_{21}, \theta^{XY}_{12}, -\theta+\theta^{XY}_{22}, \\ 
	&-\theta+\theta^{YZ}_{11}, -\theta+\theta^{YZ}_{21}, \theta^{YZ}_{12}, \theta^{YZ}_{22}, \theta^{XZ}_{11}, -\theta+\theta^{XZ}_{21}, \theta^{XZ}_{12}, \theta^{XZ}_{22}).
	\end{align*}
	The elements of $\boldsymbol{\theta}'$ determine that 21 out of 27 cell means are estimable, including cells 17 and 25, indicated in Table \ref{3by3} with asterisks. Therefore, for this model and this specified pattern of zeros, cell means $ 1,2,15,18,19,20 $ are not estimable. As these cell means are not estimable, we remove the corresponding cells from the model. This is equivalent to assuming that those observations are structural zeros. Considering $\bm\theta'$ and the 21 estimable cell means, the reduced model with three degrees of freedom is 
	$ \log \boldsymbol \mu' _{21 \times 1} = A'_{21 \times 18}\boldsymbol \theta'_{18 \times 1} $, given in the Supplementary Material.
\end{example}

\begin{example}\label{example2-gen}
	\citet{Hung:2008} performed a genome-wide association study of lung cancer by studying 500 Single Nucleotide Polymorphisms (SNP). Each SNP is categorized at levels 0, 1 and 2 to identify the number of  minor alleles.
	\citet{Papathomas:2012} selected 50 of these SNPs  via applying profile regression. We further select  five  SNPs (as representatives of uncorrelated groups of SNPs); \verb|rs7748167_C|  ($ A $),   \verb|rs4975616_G|  ($ B $),  \verb|rs6803988_T|  ($ C $),
	\verb|rs11128775_G| ($ D $),  \verb|rs9306859_A|  ($ E $).
	
	
	
	A crucial variable in this study  describes the presence or absence of cancer in each of the individuals. Adding this variable ($ F $) creates a $ 3^5 \times 2^1 $ contingency table with 486 cells. We  consider fitting a log-linear model with main effects  and  first-order interactions. This table has 298 zero cell counts and the  derivative matrix has  rank 59 with $ d=62-59=3 $. After solving the  partial differential equations for the three $ \boldsymbol{\alpha} $ vectors, the 59  estimable parameters are obtained and given in the Supplementary Material.

	Only three parameters $ \theta^{AD}_{22}, \theta^{AE}_{22}, \theta^{DE}_{22}  $ are not estimable. The estimable parameters make 360 out of 486 cell means estimable and the reduced model is, $ \log\bm\mu'_{360\times 1}=A'_{360\times 59}\bm\theta'_{59 \times 1},$
	with degrees of freedom $360-59=301$. 
	In this model, the presence of cancer has a significant positive interaction with  level $ 1 $ of variables $ A $ and $ D $ and a significant negative interaction with  level 1 of  $ C$ and $ E $ and level 2 of  $ B, C$ and $ E $.
\end{example}

\subsection{Parameter redundancy for a saturated log-linear model} 

We provide some general results on parameter redundancy for a saturated log-linear model fitted to an $l^m$  contingency table  and determine which parameters become nonestimable after observing a zero cell count. Example S1 in the Supplementary Material, illustrates the proposed approach and shows that a saturated log-linear model is always full rank when all the cell counts are  positive.

\begin{definition}
	For a saturated log-linear model, we define \textit{the parameter corresponding to the cell with count} $ y_i, i=1,\dots,n $ (according to \eqref{order}), as the one with the maximum number of variables in its superscript, within the set of all parameters in $ \log \mu_i=A_{(i)}\boldsymbol{\theta} $, where $ A_{(i)} $ is the $ i $th row of $ A $.
\end{definition}
\noindent For example, for a $3^3$ contingency table with variables $\{ X,Y,Z\} $, the parameter corresponding to observation $y_{201}$ (or $y_{12}$ according to the ordering given by \eqref{order}) is $\theta^{XZ}_{21}$. 

\begin{definition}
	For a given log-linear model parameter, \textit{parameters associated with a higher order interaction} are all those specified by including additional variables in the given parameter's superscript.
\end{definition}
\noindent For example, for the same $ 3^3 $ table, the parameters associated with a higher order interaction given  $\theta^{XZ}_{21}$, are $\theta^{XYZ}_{211}$ and $\theta^{XYZ}_{221}$.

The following theorem determines exactly which model parameters become nonestimable as a result of  a given zero observation. 

\begin{theorem}\label{onezero}
	Assume a saturated Poisson log-linear model fitted to an $l^m$  table with a single zero cell count. If $ \exists \textbf{i}, \textbf{i} \in L $ such that $ y_{\textbf{i}}=0 $,  then the parameter that corresponds to that cell, and all other parameters associated with a higher order interaction  given that parameter, are nonestimable.
\end{theorem}
\noindent  The proof by induction and examples are given in the Supplementary Material. Note that additional zero cells in the table cannot make previously nonestimable parameters estimable, as the amount of information is further reduced. Then,  the set of nonestimable parameters is at least as large as the union of the nonestimable parameters per zero cell.  The estimable parameters and linear combinations of them can be derived by solving \eqref{de}.

\subsection{The esoteric constraints}\label{esotericconstriant}
The likelihood function of parameter redundant models has a flat ridge which is occasionally orthogonal to the axes of some parameters, so  these associated parameters still have unique ML estimates \citep{Catchpole:1998}. This is when in all $ \boldsymbol{\alpha}(\boldsymbol{\theta}) $s,   the corresponding elements to these parameters are zero.  In addition, for some log-linear parameter redundant models, maximising the likelihood function  imposes one or more  extra constraints on the model parameters, due to the placement of the likelihood ridge in the parameter space. The extra constraints can make more parameters uniquely estimable compared to those specified by solving the partial differential equations in (\ref{de}).  We refer to these extra constraints as ``esoteric constraints''. Standard statistical software packages do not provide any information on these  constraints when maximising the likelihood function, so informing on them reveals the log-linear model that is, in fact, being fitted. After detecting a parameter redundant model, we can check the existence of such constraints, as explained below.

The log-likelihood function of model (\ref{loglinear model formula}) is 
$ l(\boldsymbol{\theta})= \sum_{\textbf{i}}(y_\textbf{i}\log{\mu_{\textbf{i}}(\boldsymbol{\theta})}-\mu_{\textbf{i}}(\boldsymbol{\theta}))$.  The corresponding score vector is 
$ \textbf{U}(\boldsymbol{\theta})=\left( {\partial l}/{\partial\theta_1},\cdots,{\partial l}/{\partial\theta_p}\right)^{\mathsf{T}}$, where the partial derivatives  for $s=1,\dots,p,$ are,
$$ \dfrac{\partial l}{\partial\theta_s}=\sum_{\textbf{i}}\left( \frac{y_\textbf{i}}{\mu_\textbf{i}(\boldsymbol{\theta})}-1\right) \frac{\partial \mu_\textbf{i}(\boldsymbol{\theta})}{\partial \theta_s}=\sum_\textbf{i}(y_\textbf{i}-\mu_\textbf{i}(\boldsymbol{\theta}))\frac{\partial \mu_\textbf{i}(\boldsymbol{\theta})}{\partial \theta_s}\frac{1}{\mu_\textbf{i}(\boldsymbol{\theta})}. $$
Therefore, $ \textbf{U}(\boldsymbol{\theta})=A^\mathsf{T}(\textbf{y}-\boldsymbol
{\mu}(\boldsymbol{\theta})). $ When a model is parameter redundant,  there exists at least one $ \boldsymbol{\alpha}(\boldsymbol{\theta}) $ such that $ \boldsymbol{\alpha}^{\mathsf{T}}(\boldsymbol{\theta}) D(\boldsymbol{\theta})=\textbf{0}$. If the observations are from a multinomial distribution, it follows that   $ \boldsymbol{\alpha}^{\mathsf{T}}(\boldsymbol{\theta})\textbf{U}(\boldsymbol{\theta})=0$,  which means the likelihood surface has a completely flat ridge (Theorem 2 of \citet{Catchpole:1997}). Note that,  $ \boldsymbol{\alpha}^{\mathsf{T}}(\boldsymbol{\theta})\textbf{U}(\boldsymbol{\theta})=0$ implies that the directional derivative is zero, therefore, the likelihood function is constant in  the direction of $ \boldsymbol{\alpha}(\boldsymbol{\theta}) $. This makes a ridge in the likelihood surface, which is along the curve generated by the direction field $\boldsymbol{\alpha}(\boldsymbol{\theta})$ through any point at which the likelihood is maximised.

For a Poisson log-linear model which is determined to be parameter redundant by the derivative matrix in (\ref{Dmat ct}), we set $ \boldsymbol{\alpha}^{\mathsf{T}}(\boldsymbol{\theta})\textbf{U}(\boldsymbol{\theta})=0$. The constraints that hold this equality for finite values of the model parameters, are the esoteric constraints. 
These extra constraints along with the estimable quantities in $\boldsymbol{\theta}'$,  may make more parameters estimable and permit one to obtain unique maximum likelihood estimates for parameters that otherwise would not have been estimable. Also, reducing the parameter space according to the esoteric constraints and therefore removing the flat ridge, can make it possible to uniquely maximise the likelihood. If $ \boldsymbol{\alpha}^{\mathsf{T}}(\boldsymbol{\theta})\textbf{U}(\boldsymbol{\theta})$  cannot be zero with finite $ \theta $s  then the esoteric constraints do not exist and some of the $ \theta $s tend to negative infinity.  These constraints do not exist for models described in Theorem~\ref{onezero} and in Examples~\ref{example1} and \ref{example2-gen}. A model with an esoteric constraint is given in Example~\ref{example5}. 


\section{The existence of the maximum likelihood estimator for log-linear models}\label{sec:exstMLE}

The methods summarized in this section will be referred to as the EMLE approach and will be used in  Examples~\ref{example4} and \ref{example5} in Section \ref{sec:comparison}.   We refer the reader to \citet{Fienberg:2006, Fienberg:2012a, Fienberg:2012b} for further background and  details.

Decomposable log-linear models \citep{Agresti:2002} have an explicit formula for $ \hat{\mu}_{\textbf{i}} $. For these models, positivity of minimal sufficient statistics  is a necessary and sufficient condition for the existence of the MLE of $\boldsymbol\mu $ \citep{Agresti:2002}.  For non-decomposable models, $ \hat{\mu}_{\textbf{i}} $ does not have a closed form and it  is  calculated only by iterative methods. In this case, positivity of sufficient table marginals is  still  necessary  for the existence of the estimator  but it is no longer a sufficient condition.

A condition  for the existence of the MLE of  $ \textbf{m} $ in a hierarchical log-linear model, regardless of the presence of positive or zero table marginals, was provided by \citet{Haberman:1973}.  Assume $ \mathcal{M} $ is a $ p $-dimensional linear manifold contained in  $ \mathcal{R}^{|L|} $, and 
\begin{equation}\label{habcon}
\mathcal{M}^\perp =\left\lbrace \textbf{x} \in \mathcal{R}^{|L|}: (\textbf{x},\textbf{m})=\textbf{x}^\mathsf{T}\textbf{m}=0, \forall \textbf{m} \in\mathcal{M} \right\rbrace.
\end{equation}
Then, Theorem 3.2 of \citet{Haberman:1973} states that a necessary and sufficient condition that the MLE  $ \hat{\textbf{m}} $ of $ \textbf{m} $ exists is that there is a $ \boldsymbol\delta \in \mathcal{M}^\perp $ such that $ y_\textbf{i}+\delta_\textbf{i} > 0 $ for every $ \textbf{i} \in L $. 
Here, $ \boldsymbol\mu $ in $ \textbf{m}=\log\bm\mu$ is assumed to be  positive. The theorem specifies, for any pattern of zeros in the table, whether the MLE of the cell means exists or not. In the extended maximum likelihood estimate case, a cell mean estimate could be  $ \hat{\mu}_{\textbf{i}}=0 $, but its log transformation is not defined and then estimates of  some corresponding $\theta$ parameters  tend to infinity \citep{Haberman:1974}.

A polyhedral version of  Haberman's necessary and sufficient condition  states that under any sampling design,  the MLE of $ \textbf{m} $ exists if and only if the vector of observed marginals, $ \textbf{t}=A^\mathsf{T}\textbf{y} $, lies in the relative interior of the marginal of the polyhedral cone \citep{Eriksson:2006}. The polyhedral cone,  generated by spanning columns of $ A $ with rank $ p $, is defined as, 
\begin{equation}\label{polycon}
C_A=\{\textbf{t}:\textbf{t}=A^\mathsf{T}\textbf{y}, \textbf{y} \in \mathcal{R}_{\geqslant 0}^{|L|}\}. 
\end{equation}
The MLE does not exist if and only if the vector of marginals lies on a facet or a facial set of the marginal cone \citep{Fienberg:2006}. In other words, the estimator does not exist if and only if the vector of marginals belongs to the relative interior of some proper face, $ F $, of the marginal cone. 
A face of the marginal cone is defined as a set, 
$ F =\{\textbf{t} \in C_A: (\textbf{t},\boldsymbol\zeta)=0\}$, for some $ \boldsymbol\zeta\in \mathcal{R}^p $, such that $ (\textbf{t},\boldsymbol\zeta)\geqslant 0 $ for all $ \textbf{t} \in C_{A} $, with $ (\textbf{t},\boldsymbol\zeta) $ representing the inner product.  The facial set $ \mathcal{F} $ is a set of cell indices of the rows of $ A $ whose conic hull is precisely $ F $. For any design matrix $ A $ for $ \mathcal{M}$, $ \mathcal{F} \subseteq L $ is a facial set of $ F $ if there exists some $ \boldsymbol\zeta \in \mathcal{R}^p $ such that,
\begin{align}\label{facialset}
(A_{(i)},\boldsymbol\zeta)&=0, \qquad \text{if} \quad i \in  \mathcal{F},  \\
(A_{(i)},\boldsymbol\zeta)&>0, \qquad  \text{if} \quad  i \in \mathcal{F}^c, \notag
\end{align}
where $ \mathcal{F}^c=L-\mathcal{F} $ is the co-facial set of $ F $ \citep{Fienberg:2012a}.
If such $ \boldsymbol\zeta $ and $ \mathcal{F}$ exist, the MLE does not exist and only the cell means corresponding to members of $ \mathcal{F}$ are estimable. The  nonestimable cells in $ \mathcal{F}^c $ are treated as structural zeros and are omitted from the model.  An estimable subset of  model parameters could be determined by finding 
$ A_\mathcal{F} $, the matrix whose rows are the ones from $ {A} $ with coordinates  in $ \mathcal{F} $. $ A_\mathcal{F} $ which is a $ |\mathcal{F}| \times p $ design matrix with rank $ p_F $, is then reduced to full rank $ A^*_\mathcal{F} $ with dimensions $ |\mathcal{F}| \times p_F $. By implementing this reduced design matrix, the log-likelihood function  is strictly concave with a unique maximiser. Then the extended MLE is, 
$$ \hat{\boldsymbol\theta}^e=\text{argmax}_{\boldsymbol\theta \in \mathcal{R}^{p_F}} l_\mathcal{F}(\boldsymbol\theta)=\text{argmax}_{\boldsymbol\theta \in \mathcal{R}^{p_F}} \textbf{t}_F^\mathsf{T}\boldsymbol\theta-\textbf{1}^\mathsf{T}\exp(A_\mathcal{F}^* \boldsymbol\theta),  $$
in which $ \textbf{t}_F=(A_\mathcal{F}^*)^\mathsf{T}\textbf{y}_\mathcal{F} $ and the extended
MLE of the cell mean vector is $ \hat{\textbf{m}}^e=\exp(A^*_\mathcal{F}\hat{\boldsymbol\theta}^e) $ \citep{Fienberg:2012b}.

Another way to define the facial set is by considering sub-matrices $ A_+ $ and $ A_0 $ obtained from $ A $. They are made by the rows of $ A $ indexed by $ L_+=\{{i}:y_{{i}}\neq 0\} $ and $ L_0=\{{i}:y_{{i}}= 0\} $ respectively. The vector of marginals belongs to the relative interior of some proper face of the marginal cone if and only if $ \mathcal{F}^c \subseteq L_0 $. This is equivalent to the existence of a vector $ \boldsymbol\zeta $ satisfying the following three conditions \citep{Fienberg:2012b}:
\begin{align}\label{facialsetCond}
a. & \,\, A_+ \boldsymbol\zeta=\textbf{0},   \\ 
b. & \,\,  A_0 \boldsymbol\zeta \gneq \textbf{0},   \notag \\
c. & \, \,
\text{The set} \, \{i:(A \boldsymbol\zeta)_{(i)} \neq 0\}  \, \text{has maximal cardinality among all sets of } \notag \\
&   \{i:(A \textbf{x})_{(i) }\neq 0\} \,  \text{with}  \,\, A \textbf{x}\gneq 0 \text{, for $\textbf{x}$ that satisfies the first two conditions.} \notag 
\end{align}
In (\ref{facialset}) and (\ref{facialsetCond}) the inequality signs could be changed to less than zero without loss of generality. With $\gneq 0$ we describe a non-negative vector with at least one element greater than zero.  In conclusion, if $ \text{rank}(A_+)  =  \text{rank} (A)$, the MLE exists, since no vector $ \boldsymbol\zeta $ exists and $ \mathcal{F}^c = \emptyset $. If  $ \text{rank}(A_+)  <  \text{rank} (A)$, the MLE may still exist, so we should search for a facial set.

The degrees of freedom for the reduced model is $ |\mathcal{F}|-\text{rank}(A_\mathcal{F}^*) $, which is the number of estimable cell means  minus the number of estimable model parameters \citep{Fienberg:2012a}. Computational algorithms for detecting the existence of the MLE and deriving the co-facial set, by converting these methods into linear and non-linear optimisation problems, are described by  \citet{Fienberg:2012b}. 
However, those algorithms are inefficient for a model with a large number of variables \citep{Wang:2016}. The \verb|R| packages \verb|eMLEloglin| and \verb|SparseMSE| utilize the EMLE approach to fit log-linear models \citep{Chan:2019,Friedlander:2016}.


\section{Comparison of the EMLE and the parameter redundancy approaches} \label{sec:comparison}

The two approaches described in Sections \ref{sec:parred} and \ref{sec:exstMLE} can be used to check the  identifiability of a log-linear model fitted to a sparse table. 
We compare them and summarise the comparison in the following three possible cases:
\begin{itemize}
	\item[i.] Within the EMLE framework, when the co-facial set, as defined in (\ref{facialset}), is  null, then the MLE exists. This is equivalent to the parameter redundancy outcome in which the model is not parameter redundant.
	
	\item[ii.] When there are facial and co-facial sets as defined in (\ref{facialset}), the  MLE of $ \boldsymbol\mu $ does not exist and some zero cells are treated as structural zeros. In the parameter redundancy approach, this is equivalent to having $  \boldsymbol\alpha^\mathsf{T}D=\textbf{0} $ and no esoteric constraints determined by $  \boldsymbol\alpha^{\mathsf{T}} \textbf{U}(\boldsymbol{\theta})=0 $. 
	In practice, for such a model, the determinant of the information matrix and at least one of its eigenvalues are very close to zero, considering numerical approximations and rounding errors. 
	
	\item[iii.] If there is no co-facial set as described in (\ref{facialset}), then the  MLE exists. This is equivalent to the parameter redundancy outcome in which  the model is parameter redundant with at least one  esoteric constraint that allows one to uniquely estimate the model parameters.
\end{itemize}
The next theorem explains a link between the EMLE  method and the parameter redundancy approach through the score vector $ \textbf{U}(\boldsymbol{\theta}) $. 

\begin{theorem}\label{twozero}
	For a parameter redundant model, the MLE of $ \boldsymbol\mu $ does not exist if and only if one or more $\boldsymbol{\alpha}_j$ vectors, $ j=1,\dots,d$, do not satisfy $ \boldsymbol{\alpha}_j^\mathsf{T}(\boldsymbol{\theta})\textbf{U}(\boldsymbol{\theta})={0}$ for finite elements of $\boldsymbol{\theta}$.
\end{theorem}
\noindent The proof is given in the Appendix. 

Two examples are utilized here to illustrate similarities and differences between the two approaches. Example~\ref{example4} below shows a parameter redundant model without any possible additional esoteric constraints (comparison case ii).  The two reduced models found by the two approaches have a different reparametrisation of $\boldsymbol{\theta}$, although the ML estimates of the estimable cell means are identical. The parameters in the reduced model obtained by parameter redundancy have the same interpretation as in the initial model, in terms of variable interactions.  Example~\ref{example5} presents a model that is parameter redundant and its MLE does exist (comparison case iii). This model has an esoteric constraint, extracted by the parameter redundancy approach, that  makes all parameters estimable. This  approach allows us to consider two possible ways to address the model's redundancy. Reduce the model to a smaller, saturated and non-redundant one, or adopt the esoteric constraint and estimate all parameters, which is equivalent to using numerical methods such as ``iteratively reweighted least squares'' to maximise the likelihood.  

\begin{example}\label{example4}
	We  fit  model (\ref{model23}), which can be shown as $ (XY,XZ,YZ) $, to the contingency table  in Table \ref{23ex1}(a). 
	\begin{equation}\label{model23}
	\log\mu_{ijk}=\theta+\theta^X_i+\theta^Y_j+\theta^Z_k+\theta^{XY}_{ij}+\theta^{XZ}_{ik}+\theta^{YZ}_{jk}, \qquad i,j,k=\{0,1\}^2.
	\end{equation}
	According to (\ref{order}), the vector of cell counts is $ \textbf{y}^\mathsf{T} = (y_1,y_2,y_3,y_4,y_5,y_6,y_7,y_8) \\= (y_{000}, y_{100}, y_{010}, y_{110}, y_{001}, y_{101}, y_{011}, y_{111})$. The non-zero cell counts in the table are assumed to be positive. 
	The parameter vector is shown as $ \bm\theta^\mathsf{T}=(\theta,\theta^X,\theta^Y,\theta^{XY},
	\theta^Z, \theta^{XZ},\theta^{YZ}) $ as subscripts are superfluous. The model in the form  $ \log\bm\mu_{8\times 1}=A_{8\times 7}\bm\theta_{7 \times 1}$ is given in the Supplementary Material.
	
	{\small \begin{table}[t!]  
			\caption{Observations in two $ 2^3 $ contingency tables}{%
				\setlength\tabcolsep{1.8truept}
				\begin{tabular}{c c}
					\begin{minipage}{0.52\linewidth}
						\def\arraystretch{0.9}
						\begin{tabular}{|c|cc|cc|}
							\multicolumn{5}{c}{(a)} \\
							\hline
							&\multicolumn{2}{c}{{\small $ Z=0 $}} & \multicolumn{2}{c|}{{\small $ Z=1 $}}\\
							\cline{2-5}
							& {\small $ Y=0 $} & {\small $ Y=1 $} & {\small $ Y=0 $} & {\small $ Y=1 $} \\
							\hline
							{\small $ X=0 $} & $ 0 $ & $ y_3 $  & $ y_5 $ & $ y_7 $\\  
							{\small $ X=1 $} & $ y_2 $ & $ y_4 $ & $ y_6 $ & $ 0 $\\							    
							\hline
						\end{tabular}
					\end{minipage}
					\begin{minipage}{0.52\linewidth}
						\def\arraystretch{0.9}
						\begin{tabular}{|c|cc|cc|}
							\multicolumn{5}{c}{(b)} \\
							\hline
							&\multicolumn{2}{c}{{\small $ Z=0 $}} & \multicolumn{2}{c|}{{\small $ Z=1 $}}\\
							\cline{2-5}
							& {\small $ Y=0 $} & {\small $ Y=1 $} & {\small $ Y=0 $} & {\small $ Y=1 $} \\
							\hline
							{\small $ X=0 $} & $ 0 $ & $ y_3 $  & $ y_5 $ & $ y_7 $\\
							{\small $ X=1 $} & $ y_2 $ & $ 0 $ & $ y_6 $ & $ y_8 $\\
							\hline
						\end{tabular}
					\end{minipage}
			\end{tabular}}\label{23ex1}
	\end{table}}
	
	We apply the parameter redundancy approach first. The derivative matrix formed using formula (\ref{Dmat ct})  is given in the Supplementary Material and its rank is 6, indicating that $ d=1 $. 
	From (\ref{alpha}),  $\boldsymbol \alpha^\mathsf{T}=(1,-1,-1,1,-1,1,1)$ and solving  (\ref{de}) yields the estimable parameters,
	$$ \bm\theta'^{\mathsf{T}}=(\theta+\theta^X,\theta+\theta^Y,-\theta+\theta^{XY}, \theta+\theta^Z, -\theta+\theta^{XZ},-\theta+\theta^{YZ}). $$
	Therefore, all cell means but $ \mu_{000} $ (for which, $ \log\mu_{000}=\theta $) and $ \mu_{111} $  (for which, $ \log\mu_{111}=\theta+\theta^X+\theta^Y+\theta^{XY}+\theta^Z+\theta^{XZ}+\theta^{YZ} $) are estimable. No esoteric constraint exists as,
	$$ \boldsymbol{\alpha}^\mathsf{T}\textbf{U}(\bm\theta)=y_{000}+y_{111}-e^{\theta} -e^{\theta+\theta^X+\theta^Y+\theta^{XY}+\theta^Z+\theta^{XZ}+\theta^{YZ}} \neq 0, $$
	for finite $\theta$s.
	We treat $y_{000}$  and $y_{111}$ as   structural zeros and remove them from the model.  Then, we reduce the model to a saturated one with a design matrix of rank 6 in accordance with the estimable parameters $ \bm\theta' $. The reduced non-redundant model is,
	{\small $$\left[\def\arraystretch{0.55}
		\begin{array}{c}
		\log\mu_{100}\\
		\log\mu_{010} \\
		\log\mu_{110} \\
		\log\mu_{001} \\
		\log\mu_{101} \\
		\log\mu_{011}
		\end{array}
		\right]= \left[\def\arraystretch{0.55}
		\begin{array}{ccccccc}
		1 & 0 & 0 & 0 & 0 & 0  \\
		0 & 1 & 0 & 0 & 0 & 0  \\
		1 & 1 & 1 & 0 & 0 & 0  \\
		0 & 0 & 0 & 1 & 0 & 0 \\
		1 & 0 & 0 & 1 & 1 & 0 \\
		0 & 1 & 0 & 1 & 0 & 1 
		\end{array}
		\right]\left[\def\arraystretch{0.55}
		\begin{array}{c}
		\theta+\theta^X\\
		\theta+\theta^Y\\
		-\theta+\theta^{XY}\\
		\theta+\theta^Z\\
		-\theta+\theta^{XZ}\\
		-\theta+\theta^{YZ}\\
		\end{array}
		\right]. $$}
	
	Now we consider the EMLE method. Model (\ref{model23}) has no zero sufficient marginals, but positive estimates for all the cell means do not exist according to the Haberman's sufficiency and necessary condition and also the polyhedral condition.
	To reduce this model to an identifiable one, according to the polyhedral method and (\ref{facialset}), we obtain, $ \mathcal{F}=\{100,010,110,001,101,011\}, $ 
	$ \mathcal{F}^c=\{000,111\} $,  and
	$ \boldsymbol\zeta=(1,-1,-1,1,-1,1,1) $. The design matrix for the reduced model is $ A_\mathcal{F}^* $, which is a $ |\mathcal{F}|\times p_F= 6 \times 6 $ matrix and is found by using the suggested Proposition 5.1  in  \citet{Fienberg:2012b}. The final model is,
	{\small $$\left[\def\arraystretch{0.55}
		\begin{array}{c}
		\log\mu_{100}\\
		\log\mu_{010} \\
		\log\mu_{110} \\
		\log\mu_{001} \\
		\log\mu_{101} \\
		\log\mu_{011}
		\end{array}
		\right]= \left[\def\arraystretch{0.55}
		\begin{array}{ccccccc}
		1 & 1 & 0 & 0 & 0  & 0  \\
		1 & 0 & 1 & 0 & 0  & 0  \\
		1 & 1 & 1 & 1 & 0  & 0  \\
		1 & 0 & 0 & 0 & 1  & 0 \\
		1 & 1 & 0 & 0 & 1  & 1 \\
		1 & 0 & 1 & 0 & 1  & 0 
		\end{array}
		\right]\left[\def\arraystretch{0.55}
		\begin{array}{c}
		\theta\\
		\theta^X\\
		\theta^Y\\
		\theta^{XY}\\
		\theta^Z\\
		\theta^{XZ}
		\end{array}
		\right].  $$}
	The estimable cell means are the same as derived by the parameter redundancy approach (as must be the case). However, $ \theta^{YZ} $ is dropped from the model reducing it to $ (XY,XZ) $.

	In a numerical example, the ML estimates for the six estimable cell means are identical  under the two methods and log-linear model  parameter estimates are also consistent.
	Although both methods reduce the model to one with six parameters,  parameter interpretations  differ. The parameters derived by the parameter redundancy approach are the linear combinations of the ones in the initial model. However, for instance, the estimate of $ {\theta} $ in the second reduced model is not the intercept estimate for the initial model.
\end{example}

\begin{example}\label{example5}
	Consider fitting  model (\ref{model23}) to the pattern of zeros  in  Table \ref{23ex1}(b). 
	For the parameter redundancy approach, the derivative matrix  is given in the Supplementary Material and its rank is  6, thus $ d=1$. Then, $ \boldsymbol \alpha^\mathsf{T}=(1,-1,-1,0,-1,1,1) $  indicates the estimable parameters as, 
	$$ \bm\theta'^{\mathsf{T}}=(\theta+\theta^X,\theta+\theta^Y,\theta^{XY},\theta+\theta^Z,-\theta+\theta^{XZ}, -\theta+\theta^{YZ}). $$ 
	Therefore,  $ \log \mu_{000} $ and $ \log \mu_{110} $ are not estimable. The initial model is reduced to  one with a design matrix of rank 6 as,
	
	{\small $$\left[\def\arraystretch{0.55}
		\begin{array}{c}
		\log\mu_{100}\\
		\log\mu_{010} \\
		\log\mu_{001} \\
		\log\mu_{101} \\
		\log\mu_{011} \\
		\log\mu_{111} \\
		\end{array}
		\right]= \left[\def\arraystretch{0.55}
		\begin{array}{ccccccc}
		1 & 0 & 0 & 0 & 0 & 0  \\
		0 & 1 & 0 & 0 & 0 & 0  \\
		0 & 0 & 0 & 1 & 0 & 0 \\
		1 & 0 & 0 & 1 & 1 & 0  \\
		0 & 1 & 0 & 1 & 0 & 1 \\
		1 & 1 & 1 & 1 & 1 & 1 
		\end{array}
		\right]\left[\def\arraystretch{0.55}
		\begin{array}{c}
		\theta+\theta^X\\
		\theta+\theta^Y\\
		\theta^{XY}\\
		\theta+\theta^Z\\
		-\theta+\theta^{XZ}\\
		-\theta+\theta^{YZ}\\
		\end{array}
		\right]. $$}
	However, an esoteric constraint exists and it is derived by considering, 
	$$ \boldsymbol{\alpha}^\mathsf{T}\textbf{U}(\bm\theta)=y_{000}-y_{110}-e^{\theta} +e^{\theta+\theta^X+\theta^Y+\theta^{XY}}=0. $$
	This translates to $ \theta^X+\theta^Y+\theta^{XY}=0 $ or $ \log\mu_{000}=\log\mu_{110} $. Adding this constraint on  model (\ref{model23}) makes all parameters estimable.
	
	In accordance with the EMLE approach for model (\ref{model23}), we identify a $\boldsymbol\delta$ which  satisfies (\ref{habcon}), such that $ y_\textbf{i}+\delta_\textbf{i} > 0, \forall \textbf{i} \in L $. Let $0<\delta <1$, then
	$ \boldsymbol\delta=(+\delta,-\delta,-\delta,+\delta,-\delta,+\delta,+\delta,-\delta)$
	holds the necessary and sufficient condition for the existence of the estimator of $\bm{\mu}$.
	This is also confirmed by the polyhedral condition since the observed marginals lie in the relative interior of the marginal of the polyhedral cone, as vector $ \bfy=(y_1+\delta,y_2-\delta,y_3-\delta,y_4+\delta,y_5-\delta,y_6+\delta,y_7+\delta,y_8-\delta) $  satisfies (\ref{polycon}). 
	In other words,  no $ \boldsymbol\zeta $ or $ \mathcal{F}$ can  satisfy  (\ref{facialset}) or (\ref{facialsetCond}). Thus, we are able to maximise the likelihood function by numerical methods and obtain the estimates for all parameters of model (\ref{model23}). This is possible because of the esoteric constraint, which is not reported by this method but is explicit in the parameter redundancy approach.
\end{example}


\section{Discussion}\label{sec:discussion}

We propose a parameter redundancy approach for evaluating the effect of zero cell counts on the estimability of log-linear model parameters. For a parameter redundant model, we  obtain the estimable parameters and reduce the model to an identifiable one. 

We compare the parameter redundancy approach  with  a different method that focuses on the existence of the MLE for the expected cell counts of a hierarchical model. Models with non-existent MLE are parameter redundant, whilst some log-linear models are parameter redundant despite their existent MLE. The latter happens when
maximising the likelihood function which has a flat ridge, imposes hidden extra constraints on the model to make a unique MLE possible.

The EMLE method is reported by \citet{Wang:2016} to be inefficient in finding the co-facial sets when
the number of variables in the model is larger than 16. The authors propose an approximation for the cone's face to make the method work for more variables. In the parameter redundancy approach, the symbolic algebra package \verb|Maple| can be used to simultaneously solve a number of corresponding partial differential equations. However, as \verb|Maple| runs out of memory, problems arise in the  calculations when the model deficiency increases and becomes as large as 40. The occurrence of this limitation depends on the fitted model and the pattern of zeros in the table. For example, it may become more notable in applications such as large cohort studies, when observations are concentrated in a small subspace of the entire sample space.

Future research could further explore the parameter redundant models with  existent MLE. This includes further investigating properties of the esoteric constraints and goodness of fit of the model implied by them.




\vskip 14pt
\noindent {\large\bf Supplementary Material}

The online Supplementary Material contains more details of some of the examples, Example S1, and  proof of Theorem \ref{onezero} by induction.
\par

\vskip 14pt
\noindent {\large\bf Acknowledgments}

We would like to thank the referees and the journal editor for comments that improved this manuscript. 
The work of first author is supported by EPSRC PhD grants EP/J500549/1, EP/K503162/1 and EP/L505079/1.

\vskip 14pt
\noindent {\large\bf Appendix}
\begin{proof}[\textbf{Proof of Theorem \ref{twozero}}]
	Assume the MLE does not exist for a parameter redundant model. We prove by contradiction that at least one $\boldsymbol{\alpha}_{j}$ vector does not satisfy $ \boldsymbol{\alpha}_j^\mathsf{T}(\boldsymbol{\theta})\textbf{U}(\boldsymbol{\theta})=0$ for finite elements of $ \bftheta $. Suppose  that all $\boldsymbol{\alpha}_{j}$ vectors, $j=1,\ldots,d$,  satisfy 
	$ \boldsymbol{\alpha}_j^{\mathsf{T}}(\boldsymbol{\theta})\textbf{U}(\boldsymbol{\theta})=0$ for finite elements of $ \bftheta $. 
	We know $ \textbf{U}({\boldsymbol{\theta}})={A}^{\mathsf{T}} (\textbf{y}-\bfmu(\bftheta))  $. Then,
	\begin{eqnarray}
	&\boldsymbol{\alpha}_j^{\mathsf{T}}(\boldsymbol{\theta})\textbf{U}(\boldsymbol{\theta}) = 0 \nonumber \\
	&\boldsymbol{\alpha}^{\mathsf{T}}_j {A}^{\mathsf{T}} (\textbf{y}-\bfmu(\bftheta))= 0, \nonumber \\
	&\boldsymbol{\alpha}_j^\mathsf{T} A_{+}^\mathsf{T} (\bfy-\bfmu(\bftheta))_{+} 
	+ \boldsymbol{\alpha}_j^\mathsf{T} A_{0}^\mathsf{T} (\bfy-\bfmu(\bftheta))_{0}=0, \nonumber
	\end{eqnarray}
	where $(\bfy-\bfmu(\bftheta))_{+}$ denotes a vector with the elements of $(\bfy-\bfmu(\bftheta))$ that correspond to the rows in $A_{+}$, and $(\bfy-\bfmu(\bftheta))_{0}$ denotes a vector with the elements of $(\bfy-\bfmu(\bftheta))$ that correspond to the rows in $A_{0}$. Now, $\boldsymbol{\alpha}_j^\mathsf{T} A_{+}^\mathsf{T} (\bfy-\bfmu(\bftheta))_{+}=0$, because  
	$\boldsymbol{\alpha}_j^\mathsf{T} A_{+}^\mathsf{T}=\bfzero$, since $\boldsymbol{\alpha}_j^\mathsf{T} D=\bfzero$. This implies that $\boldsymbol{\alpha}_j^\mathsf{T} A_{0}^\mathsf{T} (\bfy-\bfmu(\bftheta))_{0}=0$, or equivalently that $\boldsymbol{\alpha}_j^\mathsf{T} A_{0}^\mathsf{T} (-\bfmu(\bftheta))_{0}=0$. As the MLE does not exist, from (\ref{facialsetCond}), a $\bfzeta$ vector exists so that $A_0 \boldsymbol\zeta \gneq \bfzero$. However, $\bfzeta$ is also an $\boldsymbol{\alpha}$ vector, as $A_+ \bfzeta = \bfzero$.  Now suppose, without any loss of generality, that $\boldsymbol{\alpha}_{j^{'}}=\bfzeta$, $1\leqslant j^{'} \leqslant d$. Then, 
	\[
	A_0 \boldsymbol{\alpha}_{j^{'}} \gneq \bfzero \quad \Rightarrow \quad \boldsymbol{\alpha}_{j^{'}}^\mathsf{T} A_{0}^\mathsf{T} (-\bfmu(\bftheta))_{0} < 0,
	\]
	as all elements of $(-\bfmu(\bftheta))_{0}$ are non-zero and negative. Thus, this contradicts  $\boldsymbol{\alpha}_j^\mathsf{T} A_{0}^\mathsf{T} (-\bfmu(\bftheta))_{0}=0$. 
	
	To prove the converse, assume an $\boldsymbol{\alpha}_j$  vector exists, $1\leqslant j \leqslant d$, so that 
	$ \boldsymbol{\alpha}_j^\mathsf{T}(\boldsymbol{\theta})\textbf{U}(\boldsymbol{\theta})<0$ and cannot be zero for finite $ \boldsymbol{\theta} $. This implies that, 
	\begin{eqnarray*}\label{Th2nottrue}
		\boldsymbol{\alpha}_j^\mathsf{T} A_{+}^\mathsf{T} (\bfy-\bfmu(\bftheta))_{+} 
		+ \boldsymbol{\alpha}_j^\mathsf{T} A_{0}^\mathsf{T} (\bfy-\bfmu(\bftheta))_{0}<0, \nonumber \\
		\qquad \boldsymbol{\alpha}_j^\mathsf{T} A_{0}^\mathsf{T} (-\bfmu(\bftheta))_{0}<0,
	\end{eqnarray*}
	since $\boldsymbol{\alpha}_j^\mathsf{T} D=\bfzero$ means $ \boldsymbol{\alpha}_j^\mathsf{T}A_{+}^\mathsf{T}=\bfzero$. 
	Thus, $\boldsymbol{\alpha}_j^\mathsf{T} A_{0}^\mathsf{T} \gneq \bfzero$. 
	From all $\boldsymbol{\alpha}_j$'s so that $\boldsymbol{\alpha}_j^\mathsf{T} A_{0}^\mathsf{T} \gneq \bfzero$, we choose the $\boldsymbol{\alpha}_{j^{'}}$ that corresponds to the set $\{i:(A \textbf{x})_{(i) }\neq 0\}$ with maximal cardinality. Then, 
	$\boldsymbol{\alpha}_{j^{'}}$ satisfies the three conditions in (\ref{facialsetCond}), and the MLE does not exist. This completes the proof of Theorem \ref{twozero}. 
\end{proof}


\bibhang=1.7pc
\bibsep=2pt
\fontsize{9}{14pt plus.8pt minus .6pt}\selectfont
\renewcommand\bibname{\large \bf References}
\expandafter\ifx\csname
natexlab\endcsname\relax\def\natexlab#1{#1}\fi
\expandafter\ifx\csname url\endcsname\relax
\def\url#1{\texttt{#1}}\fi
\expandafter\ifx\csname urlprefix\endcsname\relax\def\urlprefix{URL}\fi


\end{document}


\renewcommand{\baselinestretch}{2}
	\singlespacing  
	
	
\markboth{\hfill{\footnotesize\rm S. SHARIFI FAR, M. PAPATHOMAS AND R. KING} \hfill}
{\hfill {\footnotesize\rm PARAMETER REDUNDANCY IN LOG-LINEAR MODELS} \hfill}

\renewcommand{\thefootnote}{}
$\ $\par \fontsize{12}{14pt plus.8pt minus .6pt}\selectfont

\fontsize{12}{14pt plus.8pt minus .6pt}\selectfont \vspace{0.5pc}

\Large{\centerline{\bf Supplementary Material}}
\fontsize{9}{11.5pt plus.8pt minus .6pt}\selectfont
\noindent

\setcounter{section}{0}
\setcounter{equation}{0}
\def\theequation{S\arabic{section}.\arabic{equation}}
\def\thesection{S\arabic{section}}

\fontsize{12}{14pt plus.8pt minus .6pt}\selectfont
\thispagestyle{plain}

\section{Details of paper examples}
\noindent {\bf Example 1.}
The initial log-linear model $ \log \boldsymbol \mu _{27 \times 1} = A_{27 \times 
	19}\boldsymbol \theta_{19 \times 1}, $  in the matrix form is as follows. Note that the $\mu$ indices correspond to cell counts 1 to 27 respectively.
{\scriptsize $$ 
	\left[\def\arraystretch{0.9}
	\begin{array}{c}
	\log\mu_{000} \\
	\log\mu_{100}\\
	\log\mu_{200}\\
	\log\mu_{010} \\
	\log\mu_{110} \\
	\log\mu_{210}\\
	\log\mu_{020} \\
	\log\mu_{120} \\
	\log\mu_{220}\\
	\log\mu_{001} \\
	\log\mu_{101}\\
	\log\mu_{201}\\
	\log\mu_{011} \\
	\log\mu_{111} \\
	\log\mu_{211}\\
	\log\mu_{021} \\
	\log\mu_{121} \\
	\log\mu_{221}\\
	\log\mu_{002} \\
	\log\mu_{102}\\
	\log\mu_{202}\\
	\log\mu_{012} \\
	\log\mu_{112} \\
	\log\mu_{212}\\
	\log\mu_{022} \\
	\log\mu_{122} \\
	\log\mu_{222}
	\end{array}
	\right]=  \left[\def\arraystretch{0.9}
	\begin{array}{ccccccccccccccccccc}
	1 & 0 & 0 & 0 & 0 & 0 & 0 & 0 & 0 & 0 & 0 & 0 & 0 & 0 & 0 & 0 & 0 & 0 & 0  \\
	1 & 1 & 0 & 0 & 0 & 0 & 0 & 0 & 0 & 0 & 0 & 0 & 0 & 0 & 0 & 0 & 0 & 0 & 0  \\
	1 & 0 & 1 & 0 & 0 & 0 & 0 & 0 & 0 & 0 & 0 & 0 & 0 & 0 & 0 & 0 & 0 & 0 & 0  \\
	1 & 0 & 0 & 1 & 0 & 0 & 0 & 0 & 0 & 0 & 0 & 0 & 0 & 0 & 0 & 0 & 0 & 0 & 0  \\
	1 & 1 & 0 & 1 & 0 & 0 & 0 & 1 & 0 & 0 & 0 & 0 & 0 & 0 & 0 & 0 & 0 & 0 & 0  \\
	1 & 0 & 1 & 1 & 0 & 0 & 0 & 0 & 1 & 0 & 0 & 0 & 0 & 0 & 0 & 0 & 0 & 0 & 0  \\
	1 & 0 & 0 & 0 & 1 & 0 & 0 & 0 & 0 & 0 & 0 & 0 & 0 & 0 & 0 & 0 & 0 & 0 & 0  \\
	1 & 1 & 0 & 0 & 1 & 0 & 0 & 0 & 0 & 1 & 0 & 0 & 0 & 0 & 0 & 0 & 0 & 0 & 0  \\
	1 & 0 & 1 & 0 & 1 & 0 & 0 & 0 & 0 & 0 & 1 & 0 & 0 & 0 & 0 & 0 & 0 & 0 & 0  \\
	1 & 0 & 0 & 0 & 0 & 1 & 0 & 0 & 0 & 0 & 0 & 0 & 0 & 0 & 0 & 0 & 0 & 0 & 0  \\
	1 & 1 & 0 & 0 & 0 & 1 & 0 & 0 & 0 & 0 & 0 & 0 & 0 & 0 & 0 & 1 & 0 & 0 & 0  \\
	1 & 0 & 1 & 0 & 0 & 1 & 0 & 0 & 0 & 0 & 0 & 0 & 0 & 0 & 0 & 0 & 1 & 0 & 0  \\
	1 & 0 & 0 & 1 & 0 & 1 & 0 & 0 & 0 & 0 & 0 & 1 & 0 & 0 & 0 & 0 & 0 & 0 & 0  \\
	1 & 1 & 0 & 1 & 0 & 1 & 0 & 1 & 0 & 0 & 0 & 1 & 0 & 0 & 0 & 1 & 0 & 0 & 0  \\
	1 & 0 & 1 & 1 & 0 & 1 & 0 & 0 & 1 & 0 & 0 & 1 & 0 & 0 & 0 & 0 & 1 & 0 & 0  \\
	1 & 0 & 0 & 0 & 1 & 1 & 0 & 0 & 0 & 0 & 0 & 0 & 1 & 0 & 0 & 0 & 0 & 0 & 0  \\
	1 & 1 & 0 & 0 & 1 & 1 & 0 & 0 & 0 & 1 & 0 & 0 & 1 & 0 & 0 & 1 & 0 & 0 & 0  \\
	1 & 0 & 1 & 0 & 1 & 1 & 0 & 0 & 0 & 0 & 1 & 0 & 1 & 0 & 0 & 0 & 1 & 0 & 0  \\
	1 & 0 & 0 & 0 & 0 & 0 & 1 & 0 & 0 & 0 & 0 & 0 & 0 & 0 & 0 & 0 & 0 & 0 & 0  \\
	1 & 1 & 0 & 0 & 0 & 0 & 1 & 0 & 0 & 0 & 0 & 0 & 0 & 0 & 0 & 0 & 0 & 1 & 0  \\
	1 & 0 & 1 & 0 & 0 & 0 & 1 & 0 & 0 & 0 & 0 & 0 & 0 & 0 & 0 & 0 & 0 & 0 & 1  \\
	1 & 0 & 0 & 1 & 0 & 0 & 1 & 0 & 0 & 0 & 0 & 0 & 0 & 1 & 0 & 0 & 0 & 0 & 0  \\
	1 & 1 & 0 & 1 & 0 & 0 & 1 & 1 & 0 & 0 & 0 & 0 & 0 & 1 & 0 & 0 & 0 & 1 & 0  \\
	1 & 0 & 1 & 1 & 0 & 0 & 1 & 0 & 1 & 0 & 0 & 0 & 0 & 1 & 0 & 0 & 0 & 0 & 1  \\
	1 & 0 & 0 & 0 & 1 & 0 & 1 & 0 & 0 & 0 & 0 & 0 & 0 & 0 & 1 & 0 & 0 & 0 & 0  \\
	1 & 1 & 0 & 0 & 1 & 0 & 1 & 0 & 0 & 1 & 0 & 0 & 0 & 0 & 1 & 0 & 0 & 1 & 0  \\
	1 & 0 & 1 & 0 & 1 & 0 & 1 & 0 & 0 & 0 & 1 & 0 & 0 & 0 & 1 & 0 & 0 & 0 & 1  
	\end{array}
	\right] \left[\def\arraystretch{1.2}
	\begin{array}{c}
	\theta\\
	\theta^X_1\\
	\theta^X_2\\
	\theta^Y_1\\
	\theta^Y_2\\
	\theta^Z_1\\
	\theta^Z_2\\
	\theta^{XY}_{11}\\
	\theta^{XY}_{21}\\
	\theta^{XY}_{12}\\
	\theta^{XY}_{22}\\ 
	\theta^{YZ}_{11}\\
	\theta^{YZ}_{21}\\
	\theta^{YZ}_{12}\\
	\theta^{YZ}_{22}\\
	\theta^{XZ}_{11}\\
	\theta^{XZ}_{21}\\
	\theta^{XZ}_{12}\\
	\theta^{XZ}_{22}
	\end{array}
	\right], $$}

\noindent The reduced model 	$ \log \boldsymbol \mu' _{21 \times 1} = A'_{21 \times 18}\boldsymbol \theta'_{18 \times 1} $  in the matrix form is,
{\footnotesize  $$ 
	\left[\def\arraystretch{0.9}
	\begin{array}{c}
	\log\mu_{200}\\
	\log\mu_{010} \\
	\log\mu_{110} \\
	\log\mu_{210}\\
	\log\mu_{020} \\
	\log\mu_{120} \\
	\log\mu_{220}\\
	\log\mu_{001} \\
	\log\mu_{101}\\
	\log\mu_{201}\\
	\log\mu_{011} \\
	\log\mu_{111} \\
	\log\mu_{021} \\
	\log\mu_{121} \\
	\log\mu_{202}\\
	\log\mu_{012} \\
	\log\mu_{112} \\
	\log\mu_{212}\\
	\log\mu_{022} \\
	\log\mu_{122} \\
	\log\mu_{222}
	\end{array}
	\right]=  \left[\def\arraystretch{0.9}
	\begin{array}{ccccccccccccccccccc}
	0 & 1 & 0 & 0 & 0 & 0 & 0 & 0 & 0 & 0 & 0 & 0 & 0 & 0 & 0 & 0 & 0 & 0  \\
	0 & 0 & 1 & 0 & 0 & 0 & 0 & 0 & 0 & 0 & 0 & 0 & 0 & 0 & 0 & 0 & 0 & 0  \\
	1 & 0 & 1 & 0 & 0 & 0 & 1 & 0 & 0 & 0 & 0 & 0 & 0 & 0 & 0 & 0 & 0 & 0  \\
	0 & 1 & 1 & 0 & 0 & 0 & 0 & 1 & 0 & 0 & 0 & 0 & 0 & 0 & 0 & 0 & 0 & 0  \\
	0 & 0 & 0 & 1 & 0 & 0 & 0 & 0 & 0 & 0 & 0 & 0 & 0 & 0 & 0 & 0 & 0 & 0  \\
	1 & 0 & 0 & 1 & 0 & 0 & 0 & 0 & 1 & 0 & 0 & 0 & 0 & 0 & 0 & 0 & 0 & 0  \\
	0 & 1 & 0 & 1 & 0 & 0 & 0 & 0 & 0 & 1 & 0 & 0 & 0 & 0 & 0 & 0 & 0 & 0  \\
	0 & 0 & 0 & 0 & 1 & 0 & 0 & 0 & 0 & 0 & 0 & 0 & 0 & 0 & 0 & 0 & 0 & 0  \\
	1 & 0 & 0 & 0 & 1 & 0 & 0 & 0 & 0 & 0 & 0 & 0 & 0 & 0 & 1 & 0 & 0 & 0  \\
	0 & 1 & 0 & 0 & 1 & 0 & 0 & 0 & 0 & 0 & 0 & 0 & 0 & 0 & 0 & 1 & 0 & 0  \\
	0 & 0 & 1 & 0 & 1 & 0 & 0 & 0 & 0 & 0 & 1 & 0 & 0 & 0 & 0 & 0 & 0 & 0  \\
	1 & 0 & 1 & 0 & 1 & 0 & 1 & 0 & 0 & 0 & 1 & 0 & 0 & 0 & 1 & 0 & 0 & 0  \\
	0 & 0 & 0 & 1 & 1 & 0 & 0 & 0 & 0 & 0 & 0 & 1 & 0 & 0 & 0 & 0 & 0 & 0  \\
	1 & 0 & 0 & 1 & 1 & 0 & 0 & 0 & 1 & 0 & 0 & 1 & 0 & 0 & 1 & 0 & 0 & 0  \\
	0 & 1 & 0 & 0 & 0 & 1 & 0 & 0 & 0 & 0 & 0 & 0 & 0 & 0 & 0 & 0 & 0 & 1  \\
	0 & 0 & 1 & 0 & 0 & 1 & 0 & 0 & 0 & 0 & 0 & 0 & 1 & 0 & 0 & 0 & 0 & 0  \\
	1 & 0 & 1 & 0 & 0 & 1 & 1 & 0 & 0 & 0 & 0 & 0 & 1 & 0 & 0 & 0 & 1 & 0  \\
	0 & 1 & 1 & 0 & 0 & 1 & 0 & 1 & 0 & 0 & 0 & 0 & 1 & 0 & 0 & 0 & 0 & 1  \\
	0 & 0 & 0 & 1 & 0 & 1 & 0 & 0 & 0 & 0 & 0 & 0 & 0 & 1 & 0 & 0 & 0 & 0  \\
	1 & 0 & 0 & 1 & 0 & 1 & 0 & 0 & 1 & 0 & 0 & 0 & 0 & 1 & 0 & 0 & 1 & 0  \\
	0 & 1 & 0 & 1 & 0 & 1 & 0 & 0 & 0 & 1 & 0 & 0 & 0 & 1 & 0 & 0 & 0 & 1  
	\end{array}
	\right] \left[
	\begin{array}{c}
	\theta^X_{1}\\
	\theta+\theta^X_{2}\\
	\theta+\theta^Y_{1}\\
	\theta+\theta^Y_{2}\\
	\theta+\theta^Z_{1}\\
	\theta^Z_{2}\\
	\theta^{XY}_{11}\\
	-\theta+\theta^{XY}_{21}\\
	\theta^{XY}_{12}\\
	-\theta+\theta^{XY}_{22} \\ 
	-\theta+\theta^{YZ}_{11}\\
	-\theta+\theta^{YZ}_{21}\\
	\theta^{YZ}_{12}\\
	\theta^{YZ}_{22}\\
	\theta^{XZ}_{11}\\
	-\theta+\theta^{XZ}_{21}\\
	\theta^{XZ}_{12}\\
	\theta^{XZ}_{22}.
	\end{array}
	\right]. $$}





\noindent {\bf Example 2.}
The vector of $ 59 $ estimable parameters obtained by parameter redundancy for the $ 3^5 \times 2^1 $ contingency table is,
\begin{align*}
\bm\theta^{'\mathsf{T}}=&(\theta, \theta^A_{1}, \theta^A_{2}, \theta^B_{1}, \theta^B_{2}, \theta^C_{1}, \theta^C_{2}, \theta^D_{1}, \theta^D_{2}, \theta^E_{1}, \theta^E_{2}, \theta^F_{1}, 
\theta^{AB}_{11}, \theta^{AB}_{21}, \theta^{AB}_{12}, \theta^{AB}_{22}, \theta^{AC}_{11}, \theta^{AC}_{21},\\
& \theta^{AC}_{12}, \theta^{AC}_{22}, \theta^{AD}_{11},
\theta^{AD}_{21}, \theta^{AD}_{12}, \theta^{AE}_{11}, \theta^{AE}_{21}, \theta^{AE}_{12},
\theta^{AF}_{11}, \theta^{AF}_{21},\theta^{BC}_{11}, \theta^{BC}_{21},
\theta^{BC}_{12}, \theta^{BC}_{22},  \\
&\theta^{BD}_{11}, \theta^{BD}_{21}, \theta^{BD}_{12}, \theta^{BD}_{22}, \theta^{BE}_{11}, \theta^{BE}_{21}, \theta^{BE}_{12},
\theta^{BF}_{11},  \theta^{BF}_{21}, \theta^{BE}_{22}, \theta^{CD}_{11}, \theta^{CD}_{21}, \theta^{CD}_{12},  \theta^{CD}_{22},  \\
&\theta^{CE}_{11}, \theta^{CE}_{21},  \theta^{CE}_{12},  \theta^{CE}_{22},  \theta^{CF}_{11}, \theta^{CF}_{21},  \theta^{DE}_{11}, \theta^{DE}_{21},  \theta^{DE}_{12}, \theta^{DF}_{11}, \theta^{DF}_{21}, \theta^{EF}_{11}, \theta^{EF}_{21}).
\end{align*}


\noindent {\bf Example 3.}
Model (4.11) can be written as,
{\small $$
	\left[\def\arraystretch{1}
	\begin{array}{c}
	\log\mu_{000} \\
	\log\mu_{100}\\
	\log\mu_{010} \\
	\log\mu_{110} \\
	\log\mu_{001} \\
	\log\mu_{101} \\
	\log\mu_{011} \\
	\log\mu_{111}
	\end{array}
	\right]=  \left[\def\arraystretch{1}
	\begin{array}{ccccccc}
	1 & 0 & 0 & 0 & 0 & 0 & 0  \\
	1 & 1 & 0 & 0 & 0 & 0 & 0  \\
	1 & 0 & 1 & 0 & 0 & 0 & 0  \\
	1 & 1 & 1 & 1 & 0 & 0 & 0  \\
	1 & 0 & 0 & 0 & 1 & 0 & 0 \\
	1 & 1 & 0 & 0 & 1 & 1 & 0 \\
	1 & 0 & 1 & 0 & 1 & 0 & 1 \\
	1 & 1 & 1 & 1 & 1 & 1 & 1 
	\end{array}
	\right]\left[\def\arraystretch{1}
	\begin{array}{c}
	\theta\\
	\theta^X \\
	\theta ^Y\\
	\theta^{XY} \\
	\theta^Z\\
	\theta^{XZ}\\
	\theta^{YZ} \\
	\end{array}
	\right]. $$}

\noindent The derivative matrix for contingency table in Table 2(a)   is,
{\small \begin{align*} 
	D= \left[\def\arraystretch{1}
	\begin{array}{c|cccccccc}
	& \mu_{000} & \mu_{100} & \mu_{010} & \mu_{110} & \mu_{001} & \mu_{101} & \mu_{011} & \mu_{111}  \\
	\hline
	\theta       &0&y_2&y_3&y_4&y_5&y_6&y_7&0 \\
	\theta^X     &0&y_2&0&y_4&0&y_6&0&0 \\  
	\theta^Y     &0&0&y_3&y_4&0&0&y_7&0 \\
	\theta^{XY}  &0&0&0&y_4&0&0&0&0 \\
	\theta^Z     &0&0&0&0&y_5&y_6&y_7&0 \\
	\theta^{XZ}  &0&0&0&0&0&y_6&0&0 \\
	\theta^{YZ}  &0&0&0&0&0&0&y_7&0 
	\end{array}
	\right].
	\end{align*}}

\noindent {\bf Example 4.}
The derivative matrix for contingency table in Table 2(b) is,
{\small \begin{align*} 
	D= \left[\def\arraystretch{1}
	\begin{array}{c|cccccccc}
	& \mu_{000} & \mu_{100} & \mu_{010} & \mu_{110} & \mu_{001} & \mu_{101} & \mu_{011} & \mu_{111}  \\
	\hline
	\theta       &0&y_2&y_3&0&y_5&y_6&y_7&y_8 \\
	\theta^X     &0&y_2&0&0&0&y_6&0&y_8 \\  
	\theta^Y     &0&0&y_3&0&0&0&y_7&y_8 \\
	\theta^{XY}  &0&0&0&0&0&0&0&y_8 \\
	\theta^Z     &0&0&0&0&y_5&y_6&y_7&y_8 \\
	\theta^{XZ}  &0&0&0&0&0&y_6&0&y_8 \\
	\theta^{YZ}  &0&0&0&0&0&0&y_7&y_8 
	\end{array}
	\right].
	\end{align*}}

\noindent\textbf{Example S1.}
It is known that for a log-linear model fitted to a contingency table with all positive $ y_\textbf{i} $, the log-likelihood function is strictly concave and the maximum likelihood estimates exist for all the model parameters.
Consider fitting a saturated Poisson log-linear model to an $ l^m \, ( m\geqslant 1, l\geqslant 2) $ contingency table.   The derivative matrices for a $2^1$ and a $2^2$ table are,
{\small $$ D_1=\left[\def\arraystretch{1}
	\begin{array}{c|cc}
	& \mu_{0}  & \mu_{1} \\
	\hline  
	\theta & y_1 & y_2 \\
	\theta^X &0 & y_2  
	\end{array}\right], \qquad D_2= \left[\def\arraystretch{1}
	\begin{array}{c|cccc}
	& \mu_{00} & \mu_{10} & \mu_{01} & \mu_{11}  \\
	\hline
	\theta & y_1 & y_2& y_3& y_4  \\
	\theta^X &0&y_2&0&y_4\\  
	\theta^Y &0&0&y_3&y_4\\
	\theta^{XY} &0&0&0&y_4
	\end{array}
	\right]. $$}
Even for larger tables, we can always arrange an ordering of cell means and corresponding
parameters that produces an  upper triangular $ D $ matrix in which the main diagonal elements are the cell counts, as shown in $ D_1 $ and $ D_2 $ above (and also in the proof of Theorem 1). So  when $ y_{\textbf{i}}>0,  \forall \textbf{i}\in L  $, the $ D $ matrix is always full rank, as expected, and all of the model parameters are estimable.

\section{Proof of Theorem 1}

\setcounter{equation}{0}
To prove Theorem 1, we use the induction method for two variables in two steps. First, the statement is proven to be true for an  $ l^1 $ table for all integers $ l\geqslant 2 $. Then we show that  if the statement is assumed to 
be true for an $ l^m $ table, it is also true for $ l^{m+1} $ for all integers $ l\geqslant 2 $ \citep{Earl:2017}. For simplicity, instead of  $ y_\mathsf{i} $ and $ 0 $ in the derivative matrix we write $ 1 $ and $ 0 $. This helps relate  the derivative matrix of $ m $ variables and the one with $m+1$ variables. Recall that  a zero cell turns a corresponding column to zero in the derivative matrix. To clarify the notation, without loss of generality, assume the contingency table has $ m $ variables and each of them has $ l $ levels.  
We set $D_r(\boldsymbol\theta_r)=\dfrac{d\boldsymbol\mu_r}{d\boldsymbol\theta_r}  $, in which $ \boldsymbol\mu_r $ and $ \boldsymbol\theta_r $ are the set of cell means and parameters added to the model because of adding the  $ r$th variable to the table.  Then we define 
$D_r=D_{\underline{r}}(\underline{\boldsymbol\theta_r})=\dfrac{d\underline{\boldsymbol\mu_r}}{d\underline{\boldsymbol\theta_r}}   $, as the derivative matrix for $ \underline{\boldsymbol\mu_r}=\boldsymbol\mu_1 \cup \boldsymbol\mu_2 \cup \dots \cup \boldsymbol\mu_r $ and $ \underline{\boldsymbol\theta_r}=\boldsymbol\theta_1 \cup \boldsymbol\theta_2 \cup \dots \cup \boldsymbol\theta_r $, which are union of sets of cell means and model parameters for having variables $ 1 $ to $ r $. Accordingly, $D_{r}(\underline{\boldsymbol\theta_r})=\dfrac{d{\boldsymbol\mu_r}}{d\underline{\boldsymbol\theta_r}}   $. In the tables and matrices, the $y_\mathbf{i}$'s are ordered according to (1.2) in the main paper.

Before we derive the derivative matrix and nonestimable parameters for a general case of $m=k$, we start with a simple table and gradually discover the pattern in the structure of  the derivative matrices.
For a $ 2^1 $ table, $ \boldsymbol\alpha $ and the  nonestimable parameters in presence of zero cell counts are shown here. Since only one cell count is zero,  the deficiency is one and there is one $\boldsymbol \alpha $ vector for each case. 
\begin{align*} m=1, \quad D_1&=D_{\underline{1}}(\underline{\boldsymbol\theta_1})=D_{{1}}({\boldsymbol\theta_1})=\left[\def\arraystretch{1}
\begin{array}{c|cc}
& \mu_{0}  & \mu_{1} \\
\hline  
\theta & 1 & 1 \\
\theta^X &0 & 1  
\end{array}\right],  \\  \boldsymbol\theta_1&=(\theta,\theta^X), \quad \boldsymbol\mu_1=(\mu_0,\mu_1). 
\end{align*}
{\small $$ 
	\begin{array}{ccc}
	\hline
	\text{zero cell} &  \text{$ \boldsymbol\alpha $ vector} & \text{nonestimable parameters} \\
	\hline
	y_0=0 &\boldsymbol{\alpha}_{11}=(1,-1)  & \bm\gamma_1=\{ \theta, \theta^X\} \\
	y_1=0 & \boldsymbol{\alpha}_{12}=(0,1) & \bm\gamma_2=\{\theta^X\} \\
	\hline
	\end{array}$$}\\
Those $ \boldsymbol\alpha $ vectors are actually $ {\boldsymbol\alpha}_{11}=(\alpha,-\alpha)$ and $ {\boldsymbol\alpha}_{12}=(0,\alpha) $, where $ \alpha $ could be any non-zero number but for simplification the value 1 is used.

For the model corresponding to a $ 2^2 $ table, the derivative matrix, and nonestimable parameters for setting each cell count to zero are, 
\begin{align*} m=2,\quad D_2=D_{\underline{2}}(\underline{\boldsymbol\theta_2})= \left[\def\arraystretch{1}
\begin{array}{c|cc:cc}
& \mu_{00} & \mu_{10} & \mu_{01} & \mu_{11}  \\
\hline
\theta & 1 & 1& 1& 1  \\
\theta^X &0&1&0&1\\  \hdashline
\theta^Y &0&0&1&1\\
\theta^{XY} &0&0&0&1
\end{array}
\right] \\
=\left[\def\arraystretch{0.9}
\begin{array}{cc}  
D_{\underline{1}}(\underline{\boldsymbol\theta_1}) & D_{2}({\boldsymbol{\theta}_1}) \\
\textbf{0} & D_{2}(\boldsymbol{\theta}_2) 
\end{array}\right]=\left[\def\arraystretch{0.7}
\begin{array}{cc}  
D_1 & D_{2}({\boldsymbol{\theta}_1}) \\
\textbf{0} & D_1  
\end{array}\right],
\end{align*}
\begin{align*}
\boldsymbol{\theta}_1&=(\theta,\theta^X), \qquad \quad \boldsymbol{\theta}_2=(\theta^Y,\theta^{XY}), \qquad  {\underline{\boldsymbol\theta_2}}=(\theta,\theta^X,\theta^Y,\theta^{XY}), \\
\boldsymbol{\mu}_1&=(\mu_{00},\mu_{10}), \qquad \boldsymbol{\mu}_2=(\mu_{01},\mu_{11}), \qquad {\underline{\boldsymbol\mu_2}}=(\mu_{00},\mu_{10},\mu_{01},\mu_{11}).
\end{align*}
{\small $$ 
	\begin{array}{ccc}
	\hline
	\text{zero cell} &  \text{$ \boldsymbol\alpha $ vector} & \text{nonestimable parameters} \\
	\hline
	y_{00}=y_1=0 & \boldsymbol{\alpha}_{21}=(1,-1,-1,1)=(\boldsymbol{\alpha}_{11},\boldsymbol{\alpha}_{11})& \boldsymbol\gamma_1=\{\theta, \theta^X, \theta^Y, \theta^{XY}\} \\
	y_{10}=y_2=0 & \boldsymbol{\alpha}_{22}=(0,1,0,-1)=(\boldsymbol{\alpha}_{12},\boldsymbol{\alpha}_{12}) &  \boldsymbol\gamma_2=\{\theta^X, \theta^{XY}\} \\
	y_{01}=y_3=0 & \boldsymbol{\alpha}_{23}=(0,0,1,-1)=(\textbf{0},\boldsymbol{\alpha}_{11}) & \boldsymbol\gamma_3=\{\theta^Y, \theta^{XY}\}\\
	y_{11}=y_4=0 & \boldsymbol{\alpha}_{24}=(0,0,0,1)=(\textbf{0},\boldsymbol{\alpha}_{12}) & \boldsymbol\gamma_4=\{\theta^{XY}\} \\
	\hline
	\end{array} $$}\\
The expression ${\boldsymbol\alpha}_{21}=({\boldsymbol\alpha}_{11},{\boldsymbol\alpha}_{11}) $ is true in terms of places of zero and non-zero elements which indicate estimable and nonestimable parameters. The pattern in the derivative matrices and $ \boldsymbol\alpha $ vectors holds  for increasing $ m $ and any $l$, as used in the proof below.

\begin{proof}[\textbf{Proof.}]
	
	\textit{Step one:} We prove that the statement is true for $ l^1 $ for all integers $ l\geqslant 2 $. Assume the only variable in the model is $ X $ with $ [l]=\{0,1,...,l-1\}  $ levels, therefore the saturated model includes $ l $ parameters. The derivative matrix for this model is,
	{\small $$ D_{1}=D_{\underline{1}}(\underline{\boldsymbol\theta_1})= \left[\def\arraystretch{1} \begin{array}{c|cccccc}
		& \mu_{0} & \mu_{1} & \mu_{2} & \mu_{3} & & \mu_{l-1} \\
		\hline
		\theta & 1 & 1& 1& 1& \dots & 1\\
		\theta_1^X &0&1&0& 0& \dots &0\\
		\theta_2^X &0&0&1& 0& \dots &0\\ 
		\theta_3^X &0&0&0&1& \dots &0\\ 
		\vdots & \vdots & \vdots &\vdots &\vdots &\vdots &\vdots \\
		\theta_{l-1}^{X} &0&0&0&0& \dots &1
		\end{array}\right].$$ }
	For this model, we show the $ \boldsymbol\alpha $ vectors and the  nonestimable parameters in the presence of zero cell counts. Since only one cell count is zero,  the deficiency is one and there is one $ \bm\alpha $  for each case. 
	{\small $$ \begin{array}{ccc}
		\hline
		\text{zero cell} &  \text{$ \boldsymbol\alpha $ vector} & \text{nonestimable parameters} \\
		\hline
		y_{0}=0 & \boldsymbol{\alpha}_{11}=(1,-1,-1,-1,\dots ,1) & \text{all parameters} \\
		y_{1}=0 & \boldsymbol{\alpha}_{12}=(0,1,0,0,\dots,0) &  \theta^X_1 \\
		y_{2}=0 & \boldsymbol{\alpha}_{13}=(0,0,1,0,\dots,0)& \theta^X_2\\
		y_{3}=0 & \boldsymbol{\alpha}_{14}=(0,0,0,1,\dots,0)& \theta^X_3\\
		\vdots & \vdots & \vdots \\ 
		y_{l-1}=0 & \boldsymbol{\alpha}_{1l}=(0,0,0,\dots,0,1)& \theta^X_{l-1}\\
		\hline
		\end{array}
		$$}\\
	According to the  $ \boldsymbol\alpha $ vectors, the theorem statement is true for this model.\\
	We can fix the number of variables at $ m=2 $ and show that the statement is still true for this model with any number of levels. Assume the variables in this model are $ X$ and $Y $ with $ [l]=\{0,1,...,l-1\}  $ levels, the derivative matrix for the model for this $ l^2 $ table is, 
	
	{\small 
		\begin{align*}
		D_{2}=D_{\underline{2}}(\underline{\boldsymbol\theta_2})=& \left[\def\arraystretch{0.7}
		\arraycolsep=1.4pt
		\begin{array}{c|ccccc:ccccc :c :ccccc}
		& \multicolumn{5}{c}{Y=0} & \multicolumn{5}{c}{ Y=1 } & & \multicolumn{5}{c}{ Y=l-1 } \\
		&\mu_{00}&\mu_{10}&\mu_{20}&\dots &\mu_{l-10}  &\mu_{01}&\mu_{11}&\mu_{21}&\dots &\mu_{l-11}   & \dots  &\mu_{0l-1}&\mu_{1l-1}&\mu_{2l-1}&\dots &\mu_{l-1l-1} \\
		\hline
		\theta & 1 & 1& 1& \dots & 1     & 1 & 1& 1& \dots&1  & \dots &       1 & 1& 1& \dots & 1\\
		\theta_1^X &0&1& 0& \dots &0     &0&1& 0& \dots &0    & \dots    &0&1& 0& \dots &0\\
		\theta_2^X &0&0&1&  \dots &0      &0&0&1&  \dots &0    & \dots       &0&0&1&  \dots &0\\ 
		\vdots & \vdots & \vdots  &\vdots &\vdots &\vdots      & \vdots & \vdots  &\vdots &\vdots &\vdots    & \dots    & \vdots & \vdots  &\vdots &\vdots &\vdots\\
		\theta_{l-1}^{X} &0&0&0& \dots &1     &0&0&0& \dots &1 & \dots    &0&0&0& \dots &1 \\
		\hdashline
		\theta_1^Y        & 0&0& 0& \dots &0     & 1 & 1& 1&\dots&1  & \dots     &0 & 0& 0& \dots & 0\\
		\theta_{11}^{XY}  &0&0& 0& \dots &0     &0&1& 0& \dots &0       & \dots     &0&0& 0& \dots &0\\
		\theta_{21}^{XY}  &0&0&0&  \dots &0      &0&0&1&  \dots &0      & \dots     &0&0&0&  \dots &0\\ 
		\vdots & \vdots & \vdots  &\vdots &\vdots &\vdots      & \vdots & \vdots  &\vdots &\vdots &\vdots    & \dots    & \vdots & \vdots  &\vdots &\vdots &\vdots\\
		\theta_{l-11}^{XY} &0&0&0& \dots &0     &0&0&0& \dots &1        & \dots      &0&0&0& \dots &0 \\
		\hdashline
		\vdots & &&\vdots&&&&&\vdots&&&\vdots&&&\vdots&& \\
		\hdashline
		\theta^Y_{l-1}        & 0&0& 0& \dots &0     & 0 & 0& 0& \dots&0  & \dots     &1 & 1& 1& \dots & 1\\
		\theta_{1l-1}^{XY}  &0&0& 0& \dots &0     &0&0& 0& \dots &0       & \dots     &0&1& 0& \dots &0\\
		\theta_{2l-1}^{XY}  &0&0&0&  \dots &0      &0&0&0&  \dots &0      & \dots     &0&0&1&  \dots &0\\ 
		\vdots & \vdots & \vdots  &\vdots &\vdots &\vdots       & \vdots & \vdots  &\vdots &\vdots &\vdots    & \dots    & \vdots & \vdots  &\vdots &\vdots &\vdots\\
		\theta_{l-1l-1}^{XY} &0&0&0& \dots &0     &0&0&0& \dots &0        & \dots      &0&0&0& \dots &1 
		\end{array}\right] 
		\end{align*}}
	{\small 
		\begin{flalign*}
	 =& \left[\def\arraystretch{1} 
	 \begin{array}{cccc}  
		D_{1} & D_{1} & \dots & D_{1} \\
		\textbf{0}  & D_{1} & \dots  & \textbf{0} \\
		\vdots & \vdots & \vdots & \vdots \\
		\textbf{0} & \textbf{0} & \textbf{0} & D_{1} 
	\end{array}\right]. \hspace{5cm}
	\end{flalign*}}
	The derivative matrix is upper triangular and all elements on the main diagonal are $ 1 $.  Let $ y_{\textbf{i}(0)} $ be a cell count such that its index ends with zero and $\boldsymbol{\gamma}_\textbf{i}  $ is the set including corresponding nonestimable parameters. We can order cells from $ 1 $ to $ l^m $ according to (1.2).
	Thus, in the case of having one zero cell count, the nonestimable parameters and unique $ \boldsymbol{\alpha} $ vectors are as follows which satisfy the theorem's statement.
	
	{\small
		$$
		\arraycolsep=2pt
		\def\arraystretch{1}
		\begin{array}{ccc}
		\hline
		\text{zero cell} &  \text{$ \boldsymbol\alpha $ vector} & \text{nonestimable parameters} \\
		\hline
		y_{\textbf{i}(0)}=y_{1}=0 & \boldsymbol{\alpha}_{21}=\overbrace{(\boldsymbol{\alpha}_{11},\dots,\boldsymbol{\alpha}_{11})}^{\# l}  & \boldsymbol{\gamma}_\textbf{i}=\boldsymbol{\gamma}_1=\{\text{all parameters}\}\\
		\vdots & \vdots & \vdots \\
		y_{\textbf{i}(0)}=y_{l}=0 &\boldsymbol{\alpha}_{2l}=(\boldsymbol{\alpha}_{1l},\dots,\boldsymbol{\alpha}_{1l})  &  \boldsymbol{\gamma}_\textbf{i}=\boldsymbol{\gamma}_{l}=\{\theta^X_{l-1},\theta^{XY}_{l-11},\dots, \theta^{XY}_{l-1l-1}\} \\\\
		
		y_{\textbf{i}(1)}=y_{l+1}=0 & \boldsymbol{\alpha}_{2(l+1)}=(\textbf{0},\boldsymbol{\alpha}_{11},\textbf{0},\dots,\textbf{0})  &  \boldsymbol{\gamma}_\textbf{i}=\boldsymbol{\gamma}_{l+1}=\{\theta^Y_1,\theta^{XY}_{11},\dots,\theta^{XY}_{l-11}\} \\
		\vdots & \vdots & \vdots \\
		y_{\textbf{i}(1)}=y_{l\times 2}=0  &\boldsymbol{\alpha}_{2(l\times 2)}=(\textbf{0},\boldsymbol{\alpha}_{1l},\textbf{0},\dots,\textbf{0})  &   \boldsymbol{\gamma}_\textbf{i}=\boldsymbol{\gamma}_{l\times 2}=\{\theta^{XY}_{l-11}\}  \\\\
		
		\vdots & \vdots & \vdots \\\\
		
		y_{\textbf{i}(l-1)}=y_{l^2-l+1}=0 & \boldsymbol{\alpha}_{2(l^2-l+1)}=(\textbf{0},\textbf{0},\dots,\boldsymbol{\alpha}_{11})  & \boldsymbol{\gamma}_\textbf{i}=\boldsymbol{\gamma}_{l^2-l+1}=\{\theta^Y_{l-1},\theta^{XY}_{1l-1},\dots,\theta^{XY}_{l-1l-11} \}  \\
		\vdots & \vdots & \vdots \\
		y_{\textbf{i}(l-1)}=y_{l^2}=0  &\boldsymbol{\alpha}_{2l^{2}}=(\textbf{0},\textbf{0},\dots,\boldsymbol{\alpha}_{1l})  &  \boldsymbol{\gamma}_\textbf{i}=\boldsymbol{\gamma}_{l^2}=\{\theta^{XY}_{l-1l-1} \}\\
		\hline
	\end{array}  $$ } \\

\textit{Step two:} The statement is assumed to be true for $ l^m $ when $ m=k $, we will show it is also true when $ m=k+1 $. For $ m=k $  when any of the cell counts is zero, the corresponding parameter to that cell and given that, all other parameters with a higher order interaction of the variables are assumed to be nonestimable. The derivative matrix  is, 
{ $$ D_k=D_{\underline{k}}(\underline{\boldsymbol{\theta}_k})=\left[ \begin{array}{cc}  
	D_{\underline{k-1}}(\underline{\boldsymbol{\theta}_{k-1}}) & D_{k}(\underline{\boldsymbol{\theta}_{k-1}}) \\
	{0} & D_{k}(\boldsymbol{\theta}_k) 
	\end{array}\right] = \left[ \begin{array}{cc}  
	D_{k-1} & D_{k}(\underline{\boldsymbol\theta_{k-1}}) \\
	\textbf{0} & D_{k}(\boldsymbol\theta_k) 
	\end{array}\right], $$}
in which,
{ $$ D_{k}(\boldsymbol{\theta}_{k})= {\left[\def\arraystretch{0.8}
		\begin{array}{cccc}  
		D_{k-1} & \textbf{0} & \dots & \textbf{0} \\
		\textbf{0}  & D_{k-1} & \dots  & \textbf{0} \\
		\vdots & \vdots & \vdots & \vdots \\
		\textbf{0} & \textbf{0} & \textbf{0} & D_{k-1} 
		\end{array}\right].}_{l-1 \times l-1}$$}
	
Derivative matrices are upper triangular and all elements on their main diagonals are $ 1 $.  Say $ y_{\textbf{i}(0)} $ is a cell count such that its index ends with zero. $\boldsymbol{\gamma}_\textbf{i}  $ is the set including the corresponding parameter to that cell and given that, all other parameters associated with a higher order interaction of the variables. The order of setting cell counts to zero here is the same order  used in forming the derivative matrix. 
Thus, the nonestimable parameters must be as follows (same for $ \boldsymbol{\alpha} $ vectors, because of the repetitive pattern in models and the point that in each case there is only one $ \boldsymbol{\alpha} $ vector).

{\small $$
	\arraycolsep=1pt
	\def\arraystretch{1}
	\begin{array}{ccc}
	\hline
	\text{zero cell} & \text{$ \boldsymbol\alpha $ vector} & \text{nonestimable parameters} \\
	\hline
	y_{\textbf{i}(0)}=y_{1}=0 & \boldsymbol{\alpha}_{k1}=\overbrace{(\boldsymbol{\alpha}_{k-1(1)},\dots,\boldsymbol{\alpha}_{k-1(1)})}^{\# l}  & \boldsymbol{\gamma}_\textbf{i}=\boldsymbol{\gamma}_1=\{\text{all parameters}\}\\
	\vdots & \vdots & \vdots \\
	y_{\textbf{i}(0)}=y_{l^{k-1}}=0 &\boldsymbol{\alpha}_{kl^{k-1}}=(\boldsymbol{\alpha}_{k-1(l^{k-1})},\dots,\boldsymbol{\alpha}_{k-1(l^{k-1})})  &  \boldsymbol{\gamma}_\textbf{i}=\boldsymbol{\gamma}_{l^{k-1}} \\\\
	
	y_{\textbf{i}(1)}=y_{l^{k-1}+1}=0 & \boldsymbol{\alpha}_{k(l^{k-1}+1)}=(\textbf{0},\boldsymbol{\alpha}_{k-1(1)},\textbf{0},\dots,\textbf{0})  &  \boldsymbol{\gamma}_\textbf{i}=\boldsymbol{\gamma}_{l^{k-1}+1} \\
	\vdots & \vdots & \vdots \\
	y_{\textbf{i}(1)}=y_{l^{k-1}\times 2}=0  &\boldsymbol{\alpha}_{k(l^{k-1}\times 2)}=(\textbf{0},\boldsymbol{\alpha}_{k-1(l^{k-1})},\textbf{0},\dots,\textbf{0})  &   \boldsymbol{\gamma}_\textbf{i}=\boldsymbol{\gamma}_{l^{k-1}\times 2}  \\\\
	
	\vdots & \vdots & \vdots \\\\
	
	y_{\textbf{i}(l-1)}=y_{(l^{k-1}\times l-1)+1}=0 & \boldsymbol{\alpha}_{k((l^{k-1}\times l-1)+1)}=(\textbf{0},\textbf{0},\dots,\boldsymbol{\alpha}_{k-1(1)})  & \boldsymbol{\gamma}_\textbf{i}=\boldsymbol{\gamma}_{(l^{k-1}\times l-1)+1}  \\
	\vdots & \vdots & \vdots \\
	y_{\textbf{i}(l-1)}=y_{l^{k}}=0  &\boldsymbol{\alpha}_{kl^{k}}=(\textbf{0},\textbf{0},\dots,\boldsymbol{\alpha}_{k-1(l^{k-1})})  &  \boldsymbol{\gamma}_\textbf{i}=\boldsymbol{\gamma}_{l^{k}}=\{\text{only the} \\
	& &  \text{highest order parameter}\}   \\
	\hline
\end{array}  $$}

Now the theorem statement must be proven for $m=k+1$. We have,
{ 	
$$ D_{k+1}=D_{\underline{k+1}}(\underline{\boldsymbol{\theta}_{k+1}})=\left[ \begin{array}{cc}  
D_{\underline{k}}(\underline{\boldsymbol{\theta}_k}) & D_{k+1}(\underline{\boldsymbol{\theta}_{k}}) \\
{0} & D_{k+1}(\boldsymbol{\theta}_{k+1}) 
\end{array}\right] \\
= \left[ \begin{array}{cc}  
D_{k} & D_{k+1}(\underline{\boldsymbol{\theta}_{k}}) \\
{0} & D_{k+1}(\boldsymbol{\theta}_{k+1}) 
\end{array}\right],
$$}
in which,
{ $$ D_{k+1}(\boldsymbol{\theta}_{k+1})= {\left[\def\arraystretch{0.7}
	\begin{array}{cccc}  
	D_{k} & \textbf{0} & \dots & \textbf{0} \\
	\textbf{0}  & D_{k} & \dots  & \textbf{0} \\
	\vdots & \vdots & \vdots & \vdots \\
	\textbf{0} & \textbf{0} & \textbf{0} & D_{k} 
	\end{array}\right].}_{l-1 \times l-1}$$}
So the nonestimable parameters are expected to be,
{\small 
$$
\arraycolsep=2pt
\def\arraystretch{1}
\begin{array}{cc}
\hline 
\text{zero cell}  & \text{nonestimable parameters} \\
\hline
y_{\textbf{i}(0)}=y_{1}=0 &  \boldsymbol{\gamma}_\textbf{i}=\boldsymbol{\gamma}_1=\{\text{all parameters}\}\\
\vdots &  \vdots \\
y_{\textbf{i}(0)}=y_{l^{k}}=0 &  \boldsymbol{\gamma}_\textbf{i}=\boldsymbol{\gamma}_{l^{k}} \\\\

y_{\textbf{i}(1)}=y_{l^{k}+1}=0 & \boldsymbol{\gamma}_\textbf{i}=\boldsymbol{\gamma}_{l^{k}+1} \\
\vdots &  \vdots \\
y_{\textbf{i}(1)}=y_{l^{k}\times 2}=0  &   \boldsymbol{\gamma}_\textbf{i}=\boldsymbol{\gamma}_{l^{k}\times 2}  \\\\

\vdots &  \vdots \\\\

y_{\textbf{i}(l-1)}=y_{(l^{k}\times l-1)+1}=0 &  \boldsymbol{\gamma}_\textbf{i}=\boldsymbol{\gamma}_{(l^{k}\times l -1)+1}  \\
\vdots &  \vdots \\
y_{\textbf{i}(l-1)}=y_{l^{k+1}}=0  &  \boldsymbol{\gamma}_\textbf{i}=\boldsymbol{\gamma}_{l^{k+1}}=\{\text{only the} \\
&  \text{highest order parameter}\}   \\
\hline
\end{array}  $$}
{To prove that these are nonestimable parameters, we need to obtain the corresponding $ \boldsymbol{\alpha} $ vectors.  According to the repetitive pattern of $ \boldsymbol{\alpha} $ vectors, that was observed when constructing the derivative matrices by increasing the number of variables in the  table, they are made of vectors of the previous step. Therefore the unique $ \boldsymbol{\alpha} $ vectors are,}
{\small $$
\arraycolsep=2pt\
\def\arraystretch{1}
\begin{array}{cc}
\hline 
\text{zero cell}  & \text{$ \boldsymbol\alpha $ vector} \\
\hline
y_{\textbf{i}(0)}=y_{1}=0 & \boldsymbol{\alpha}_{k+1(1)}=\overbrace{(\boldsymbol{\alpha}_{k1},\dots,\boldsymbol{\alpha}_{k1})}^{\# l} \\
\vdots & \vdots  \\
y_{\textbf{i}(0)}=y_{l^{k}}=0 &\boldsymbol{\alpha}_{k+1(l^{k})}=(\boldsymbol{\alpha}_{kl^{k}},\dots,\boldsymbol{\alpha}_{kl^{k}})  \\\\

y_{\textbf{i}(1)}=y_{l^{k}+1}=0 & \boldsymbol{\alpha}_{k+1(l^{k}+1)}=(\textbf{0},\boldsymbol{\alpha}_{k1},\textbf{0},\dots,\textbf{0})   \\
\vdots & \vdots  \\
y_{\textbf{i}(1)}=y_{l^{k}\times 2}=0   &\boldsymbol{\alpha}_{k+1(l^{k}\times 2)}=(\textbf{0},\boldsymbol{\alpha}_{kl^{k}},\textbf{0},\dots,\textbf{0})    \\\\

\vdots & \vdots  \\\\

y_{\textbf{i}(l-1)}=y_{(l^{k}\times l-1)+1}=0 & \boldsymbol{\alpha}_{k+1((l^{k}\times l-1)+1)}=(\textbf{0},\textbf{0},\dots,\boldsymbol{\alpha}_{k1})  \\
\vdots & \vdots \\
y_{\textbf{i}(l-1)}=y_{l^{k+1}}=0 &\boldsymbol{\alpha}_{k+1l^{k+1}}=(\textbf{0},\textbf{0},\dots,\boldsymbol{\alpha}_{kl^k})     \\
\hline
\end{array}  $$}

{ For the first $ \frac{1}{l} $ proportion of the cases in the previous table, having a zero cell count makes $ \boldsymbol{\alpha}=(\boldsymbol{\alpha}_{ki},\dots,\boldsymbol{\alpha}_{ki}) $. Since the theorem is assumed to be true for $ m=k $, the first  $ \boldsymbol{\alpha}_{ki} $ makes the corresponding parameter to that cell and given that, all other parameters with a higher order interaction of variables be nonestimable for the last smaller model ($ m=k $). Repeating $ \boldsymbol{\alpha}_{ki} $, $ l-1 $ times in the $ \boldsymbol{\alpha} $  vector makes some other parameters of the new model to be nonestimable, which are the same previous parameters corresponding to  all levels of the  new variable. Hence, the corresponding parameter to that cell and given that, all other parameters with a higher order interaction of the variables are nonestimable.

For the rest of the $ \frac{1}{l} $ parts of the cases, having a zero cell count makes an $ \boldsymbol{\alpha}_{ki} $  appear in the vector. This $ \boldsymbol{\alpha}_{ki} $ makes the corresponding parameter to that cell and given that, all other parameters with a higher order interaction of the variables be nonestimable for the last smaller model, but as it appeared after one or more vectors of zeroes here, those parameters will have the higher levels of the new variable in their superscript and subscript. Hence, the corresponding parameter to that cell and given that, all other parameters with a higher order interaction of the variables are nonestimable. Therefore the statement is true for $ m=k+1 $. }
\end{proof}
